\documentclass[
amsmath,amssymb,
prd,nofootinbib,floatfix,12pt,
]{revtex4}

\usepackage{amsmath,amssymb,amsfonts}
\usepackage[dvips]{color}
\usepackage{xcolor}
\usepackage{dsfont}
\usepackage{mathrsfs}
\usepackage{float}
\usepackage{graphicx}
\usepackage{graphics}
\usepackage{setspace}
\usepackage{pdfpages}
\usepackage{lmodern}

\usepackage{hyperref}

\hypersetup{colorlinks,citecolor=blue,urlcolor=blue, linkcolor=black}
\newcommand\beq{\begin{eqnarray}}
\newcommand\eeq{\end{eqnarray}}

\allowdisplaybreaks
\newcommand\missET{E_T^{\rm miss}}

\def\lsim{\mathrel{\rlap{\lower4pt\hbox{$\sim$}}
    \raise1pt\hbox{$<$}}}                
\def\gsim{\mathrel{\rlap{\lower4pt\hbox{$\sim$}}
    \raise1pt\hbox{$>$}}}

\begin{document}
\renewcommand{\theequation}{\arabic{section}.\arabic{equation}}
\renewcommand{\thefigure}{\arabic{section}.\arabic{figure}}
\renewcommand{\thetable}{\arabic{section}.\arabic{table}}

\title{\large Chasing higgsino dark matter at colliders in the neutrino fog era}

\author{Prudhvi N. Bhattiprolu$^{\ast}$,
        Stephen P. Martin$^{\dag}$,
        James D. Wells$^{\ddag}$
}

\affiliation{\it \baselineskip=18pt
  $^{\ast}$ INFN, Sezione di Firenze, Via G. Sansone 1, 50019 Sesto Fiorentino, Italy\\
  $^{\dag}$ Department of Physics, Northern Illinois University, DeKalb IL 60115, USA\\
  $^{\ddag}$ Leinweber Institute for Theoretical Physics, University of Michigan, Ann Arbor MI 48109, USA
}

\begin{abstract}\normalsize\baselineskip=17pt
Higgsinos can be the lightest supersymmetric particles, allowing for either a full or partial dark matter interpretation, with the correct thermal freeze-out abundance obtained for masses near 1.1 TeV. Dark matter direct detection experimental results, now rapidly approaching the neutrino fog, imposes increasingly stringent requirements on higgsino purity. We begin by summarizing the purity constraints implied by the  current strong limits from the LUX-ZEPLIN experiment in 2024, presenting them as lower bounds on gaugino masses in scenarios with decoupled sfermions and heavy Higgs bosons. We further quantify how these constraints will evolve as direct detection approaches various neutrino fog discovery and exclusion definitions and future exclusion projections. Finally, given that nearly pure higgsinos remain notoriously challenging to probe directly at colliders, we explore complementary signatures in which higgsinos are produced from the decays of heavier superpartners, where additional leptons and jets can be used for triggering. In particular, we advocate for searches of stop and wino pairs decaying directly to higgsinos as a promising means to probe higgsino dark matter well into the neutrino fog era.
\end{abstract}

\maketitle
\baselineskip=17pt

\begingroup \setstretch{0.98} \tableofcontents \endgroup
\baselineskip=17pt

\newpage

\section{Introduction\label{sec:intro}}
\setcounter{equation}{0}
\setcounter{figure}{0}
\setcounter{table}{0} 
\setcounter{footnote}{1}

The presence of dark matter is strongly implied by various astrophysical observations, but its particle nature remains elusive.
A weakly interacting massive particle (WIMP) with weak-scale mass and couplings can roughly accommodate the correct thermal relic abundance of dark matter observed today, $\Omega_\text{DM} h^2 = 0.12$ \cite{Planck:2018vyg}.
Softly broken supersymmetry (for reviews, see refs.~\cite{Martin:1997ns,Dreiner:2023yus}), often motivated to explain the origin and stability of the weak scale, can naturally feature such a dark matter candidate.

Neutralino dark matter within the Minimal Supersymmetric Standard Model (MSSM) is one of the quintessential examples of the WIMP paradigm \cite{Arkani-Hamed:2004ymt,Giudice:2004tc,Pierce:2004mk,Profumo:2004at,Cheung:2012qy}. In this framework, the lightest neutralino is the lightest supersymmetric particle (LSP), stable under the assumption of $R$-parity, and undergoes standard thermal freeze-out to generate the required dark matter relic abundance.
While the parameter space in such a scenario--or for supersymmetric dark matter in general--is constrained by various direct and indirect dark matter searches, as well as collider probes for superpartners, a neutralino LSP that is nearly a pure higgsino remains an attractive possibility.
Moreover, with wino dark matter now excluded by Fermi gamma-ray observations \cite{Safdi:2025sfs}, higgsino dark matter is perhaps one of the last minimal and viable WIMP scenarios that remain well-motivated within supersymmetry and beyond \cite{Cirelli:2007xd,Bottaro:2022one}.

In order for supersymmetry to address the hierarchy between the weak and Planck scales without excessive fine-tuning, the higgsino mass parameter $\mu$ is generally expected to be not far above the weak scale. The discovery of the 125 GeV Higgs boson, together with null results from direct superpartner searches at the Large Hadron Collider (LHC) and indirect constraints from flavor and CP-violating observables, is consistent with electroweakinos being somewhat heavier than the weak scale, while the gluino, scalar superpartners, and additional Higgs bosons are significantly heavier and effectively decoupled \cite{Wells:2003tf,Arkani-Hamed:2004zhs,Wells:2004di,Arvanitaki:2012ps}.
Therefore, higgsinos can easily be the lightest supersymmetric particles, allowing for either a full or partial dark matter interpretation.
Note that since the masses of gauginos, sfermions, and heavy Higgs bosons can arise entirely from soft supersymmetry-breaking terms, they are not fixed by electroweak symmetry breaking and need not be light. A heavy superpartner spectrum is also favored by the precision unification of gauge couplings \cite{Bhattiprolu:2023lfh}.

In this paper, we quantify present and future bounds on scenarios where a mostly higgsino LSP constitutes some or all of the thermal dark matter abundance, and examine their implications for LHC searches. Nearly degenerate higgsinos are notoriously challenging to probe at the LHC, and current indirect dark matter searches are not constraining. We therefore begin by summarizing the acute higgsino purity constraints emerging from
the rapidly evolving dark matter direct detection program \cite{PandaX-4T:2021bab,XENON:2023cxc,LZ:2022lsv,PandaX:2024qfu,LZ:2024zvo,XLZD:2024nsu,Baudis:2024jnk,PANDA-X:2024dlo},
which will ultimately encounter the neutrino fog \cite{Billard:2013qya}.
Specifically, we consider the current strong limits from the LUX-ZEPLIN experiment in 2024 (LZ2024) \cite{LZ:2024zvo}, future projections for XLZD \cite{XLZD:2024nsu,Baudis:2024jnk} with 200 and 1000 tonne-years of exposure, PandaX-xT \cite{PANDA-X:2024dlo} with 200 tonne-years, and several definitions of the neutrino fog. We present these purity constraints as lower bounds on the gaugino mass parameters, with decoupled sfermions and heavy Higgs bosons, in two well-motivated gaugino mass hierarchies: (i) models with gaugino mass unification, where the bino is the lightest gaugino, and (ii) anomaly-mediated supersymmetry breaking (AMSB) models \cite{Randall:1998uk,Giudice:1998xp}, where the wino is the lightest gaugino.

To overcome the challenges of detecting higgsinos through direct searches that arise from their low pair-production cross sections and small mass splittings, we explore the prospects of detecting them via the decays of other superpartners, where additional leptons and jets can be used for triggering. In models with gaugino mass unification, the purity constraints, as we will see below, can imply that the gluino lies well beyond the LHC’s reach. This leaves the lightest top squark a promising search target if its mass is driven down by significant level repulsion in the stop sector. In contrast, in AMSB models, the wino can be a viable LHC target. Beyond ongoing direct higgsino searches, we advocate for searches of stop pairs and wino pairs decaying directly to higgsinos, generalized to a broader range of higgsino mass splittings, as a means to probe a nearly pure higgsino LSP consistent with a full or partial dark matter interpretation.

\section{Higgsino dark matter\label{sec:HDM}}
\setcounter{equation}{0}
\setcounter{figure}{0}
\setcounter{table}{0} 
\setcounter{footnote}{1}

The possibility of a nearly pure higgsino LSP constituting some or all of dark matter, besides being phenomenologically well-motivated as argued in the previous section, also provides a simple framework with only a few parameters: the higgsino mass parameter $\mu$, the bino mass $M_1$, the wino mass $M_2$, and the ratio of Higgs vacuum expectation values $\tan \beta$. We take $\mu$, $M_1$, $M_2$ to be real, to easily avoid constraints from the electron electric dipole moment and CP violating constraints in general. For simplicity, we also take $M_1$ and $M_2$ to be positive.

We consider the case where the lightest neutralino LSP $\tilde{N}_1$ is mostly higgsino and accounts for some or all of the dark matter. The other higgsino-like states--a chargino $\tilde C_1$ and another neutralino $\tilde{N}_2$--are slightly heavier than $m_{\tilde{N}_1}$, with small and positive mass splittings
\beq
    \Delta M_+ = m_{\tilde{C}_1} - m_{\tilde{N}_1},\quad
    \Delta M_0 = m_{\tilde{N}_2} - m_{\tilde{N}_1},
\eeq
with all three higgsino-like states nearly degenerate around the higgsino mass parameter $\mu$, and
the gaugino-like neutralinos $\tilde{N}_3, \tilde{N}_4$ and a wino-like chargino $\tilde C_2$ heavier than the higgsino-like states. The gluino and scalars (squarks, sfermions, and heavy Higgs bosons) are taken to be essentially decoupled. For all numerical results, we fix the lightest Higgs boson mass to 125.1 GeV.

The splitting $\Delta M_+$ receives a positive electroweak radiative contribution, at most 355 MeV, which is especially important in the pure higgsino limit with decoupled gauginos \cite{Thomas:1998wy}.
Higgsino dark matter would be excluded by inelastic $Z$-exchange scattering of $\tilde{N}_1$ to $\tilde{N}_2$ in direct detection experiments if the higgsinos were pseudo-Dirac with $\Delta M_0 \lesssim 200$ keV \cite{Tucker-Smith:2001myb,Graham:2024syw}, which would require extremely heavy gauginos at more than a few tens of PeV. However, this is not the case in natural supersymmetry with lighter gauginos, and moreover, with the higgsino-gaugino mass mixing terms being proportional to the electroweak scale.

Coannihilation of higgsino-like states plays an important role in setting the relic abundance via standard thermal freeze-out \cite{Griest:1990kh}. In order to obtain the observed dark matter thermal abundance of $\Omega_\text{LSP} h^2 = 0.12$ with a nearly pure higgsino LSP, it is well known that the LSP mass should be around 1.1 TeV. Consequently, the thermal abundance of LSP $\Omega_\text{LSP} h^2 < 0.12$ if $m_{\tilde{N}_1}$ is less than this characteristic value. The Large Electron-Positron (LEP) collider bound on charginos implies that the LSP mass be greater than about 100 GeV which can still account for percent level dark-matter density.
In the case of a partial dark matter interpretation, other species such as axions (see, e.g., refs.~\cite{Bae:2013bva,Bae:2014yta,Bae:2017hlp,Baer:2019uom}) can account for the rest of the dark matter density, suppressing direct detection rates by a factor of $\Omega_\text{LSP} h^2/0.12$.

Experimentally, as commented earlier, nearly degenerate higgsinos are very challenging to probe at the LHC, with current searches limited to $m_{\tilde{N}_1} \lesssim 200$ GeV.
Indirect detection of higgsino annihilations in the galactic center through gamma-ray telescopes is not currently constraining, but has good discovery prospects \cite{Rinchiuso:2020skh,Rodd:2024qsi,Abe:2025lci,Baumgart:2025dov}
for a 1.1 TeV thermal higgsino at the future Cherenkov Telescope Array Observatory \cite{CTAConsortium:2017dvg} and Southern Wide-field Gamma-ray Observatory \cite{Albert:2019afb}, assuming optimistic dark matter density profiles.
Dark matter direct detection experiments impose purity constraints on higgsinos, which can be translated into lower bounds on gaugino mass parameters and upper bounds on the higgsino mass splittings $\Delta M_0$ and $\Delta M_+$, with decoupled scalars, as discussed in refs.~\cite{Martin:2024pxx,Martin:2024ytt}.

To interpret the purity constraints implied by direct detection searches, we consider two well-motivated gaugino mass hierarchies. One scenario is gaugino mass unification, motivated by the apparent unification of gauge couplings in the MSSM near $2 \times 10^{16}$ GeV, the grand unified theory (GUT) scale. In this framework, the gaugino masses $M_1$, $M_2$, and $M_3$ are assumed to unify at the GUT scale. Renormalization group running renders the bino the lightest gaugino, so that the lightest gaugino-like neutralino $\tilde{N}_3$ is predominantly bino-like. A second scenario is realized in AMSB models, where a distinctive gaugino mass hierarchy arises from the superconformal anomaly and is determined by renormalization group quantities. In these models, the wino is the lightest gaugino, making $\tilde{N}_3$ predominantly wino-like.
The higgsino purity constraints implied by dark matter direct detection bounds can therefore be translated to lower bounds on the bino (wino) mass in models with GUT-normalized (special anomaly-mediated) gaugino mass hierarchy.
We quantify these bounds in the next section.
For concreteness, we take $M_2 = 1.8 M_1$ in models with gaugino mass unification and $M_1 = 3.2 M_2$ in AMSB models.

We employed \texttt{micrOMEGAs v6.0} \cite{Belanger:2001fz,Belanger:2004yn,Belanger:2020gnr,Alguero:2023zol} to calculate relic densities of thermal higgsino LSPs and their spin-independent (SI) and spin-dependent (SD) direct detection cross sections. We made use of \texttt{SuSpect v2.41} \cite{Djouadi:2002ze}, interfaced with \texttt{micrOMEGAs}, to compute the couplings and physical masses at one-loop order, and we computed the gluino pole mass at one-loop using \texttt{SuSpect v3.1.1} \cite{Kneur:2022vwt}. Finally, \texttt{SOFTSUSY v4.1.20} \cite{Allanach:2001kg}  was used to calculate $\Delta M_+$ from underlying model parameters, which is needed to determine the proper decay length of the higgsino-like chargino in section~\ref{sec:HDMatcolliders}, as it agrees well with ref.~\cite{Thomas:1998wy} in the decoupled gaugino limit.

\section{Higgsino purity constraints in the neutrino fog era\label{sec:DMdetecion}}
\setcounter{equation}{0}
\setcounter{figure}{0}
\setcounter{table}{0}
\setcounter{footnote}{1}

At present, the most stringent direct detection limits on higgsino dark matter come from LZ2024.
As the sensitivities of dark matter direct detection experiments improve, they will become limited by the background from
coherent elastic scattering of astrophysical neutrinos from nuclei. For the mass range between 100 and 1200 GeV relevant for higgsinos, the main background source is from atmospheric neutrinos initiated by cosmic rays.
More generally, this limitation is often known as the ``neutrino floor" or ``neutrino fog".  
In ref.~\cite{OHare:2021utq}, a definition of neutrino fog level was proposed in the following way. Consider a discovery or exclusion criterion for the WIMP-nucleon cross-section $\sigma$ at some significance $Z$ as a function of the observed number of background events $N$. Now define an index $n$ by
\beq
1/n = - \frac{N}{\sigma} \frac{d\sigma}{dN}. 
\label{eq:defnnufog}
\eeq
Then cross-sections such that $n>2$ are considered to be below the neutrino floor, or in the neutrino fog. This corresponds to the limit scaling worse than the Poisson case $\sigma\propto 1/\sqrt{N}$. 

In the examples of ref.~\cite{OHare:2021utq}, the cross-section appearing in eq.~(\ref{eq:defnnufog}) was taken to be the one for discovery with significance $Z=3$. However, other choices are possible, and can give quite different results. Another possibility, seen in some experimental projections, is to take $\sigma$ needed for an exclusion at 90\% CL, which results in a lower neutrino fog cross-section definition. 
An obvious alternative choice would be to use the $\sigma$ needed for an expected discovery significance $Z=5$, 
which results in a higher neutrino fog cross-section definition. Another, perhaps more directly practical, way of quantifying the limitations due to the neutrino background is to estimate the projected 90\% CL excluded cross-section for a particular proposed experiment.

\begin{figure}[t]
\mbox{
\includegraphics[width=0.515\linewidth]{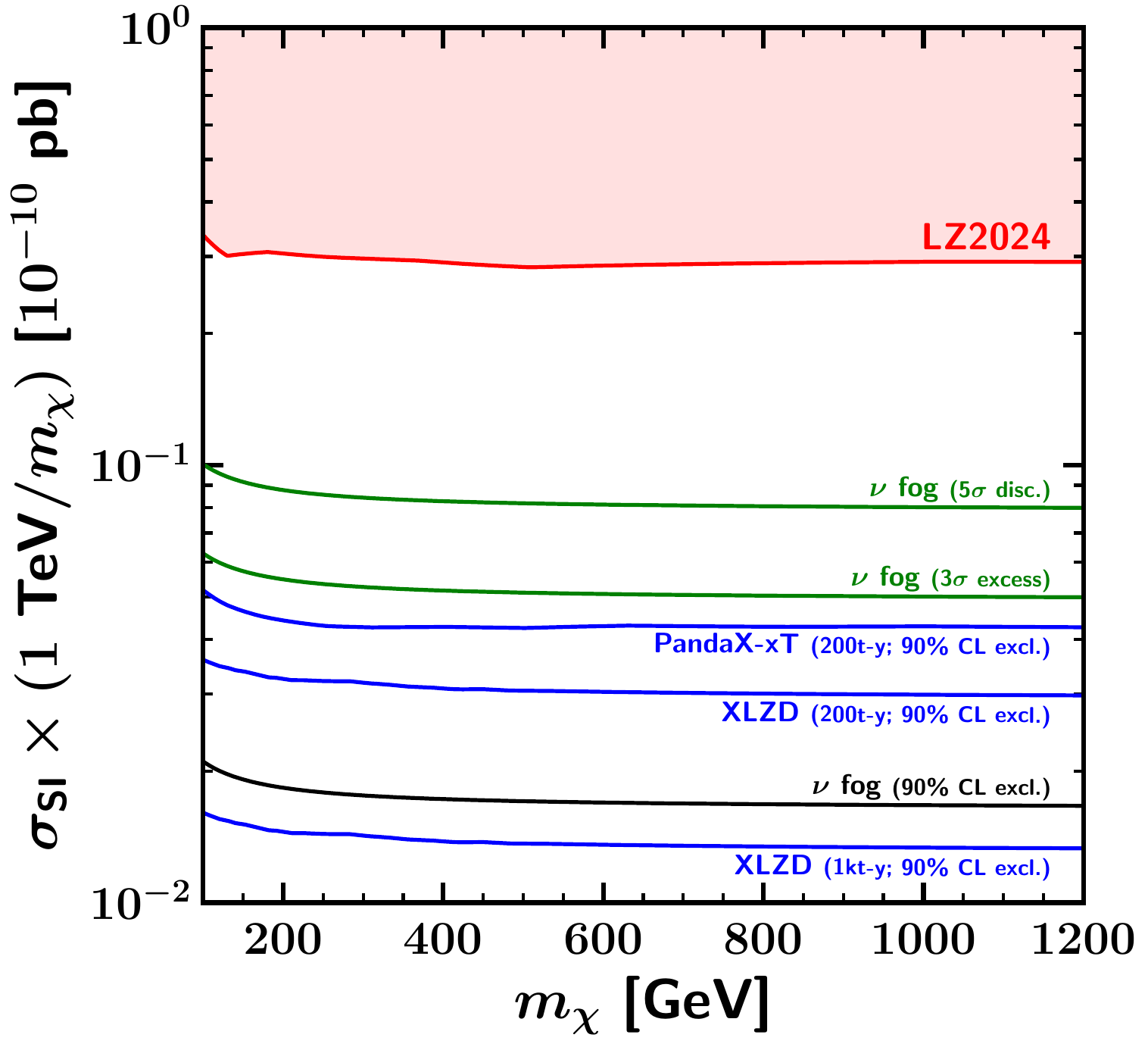}
\includegraphics[width=0.515\linewidth]{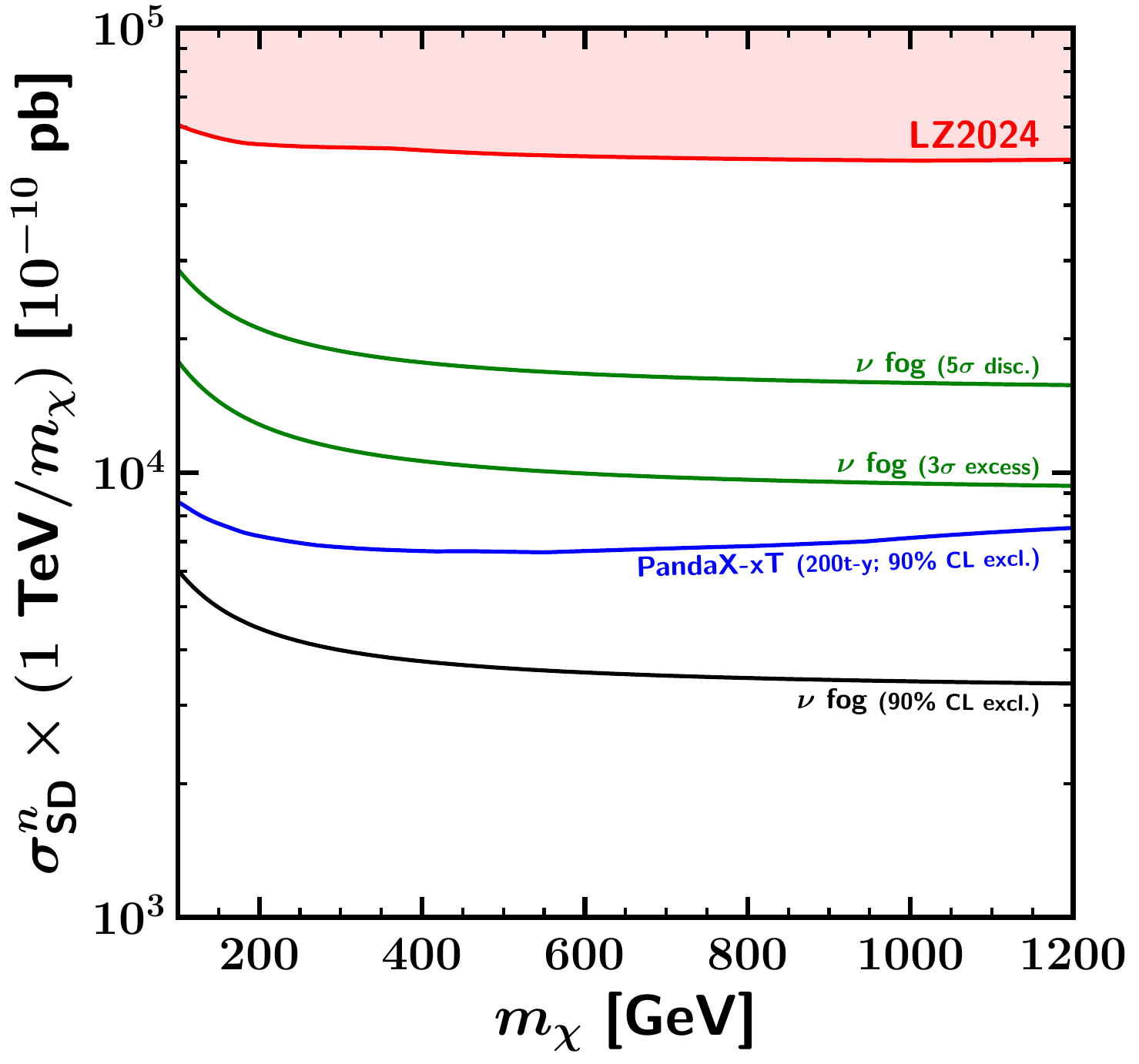}
}
\caption{\label{fig:foglevels}
Comparison of different neutrino fog discovery and exclusion definitions for the spin-independent nucleon (SI, left panel) and spin-dependent neutron (SDn, right panel) cross-sections. The vertical axes show the cross-sections scaled by (1 TeV$/m_{\chi}$), so that present and future limits are close to horizontal lines. Results are shown as a function of the WIMP mass $m_{\chi}$ in the range relevant for thermal dark matter higgsinos. The upper shaded region in each figure is the current LZ2024 90\% CL exclusion. In the left panel, the other lines are, from top to bottom: the $5\sigma$ discovery and $3\sigma$ excess neutrino fog levels,
the projected PandaX-xT and XLZD 90\% CL limits with 200 tonne-years, the 90\% CL exclusion neutrino fog,
and the XLZD 90\% CL limit with 1000 tonne-years. The right panel shows
the $5\sigma$ discovery and $3\sigma$ excess neutrino fog lines, the projected PandaX-xT 90\% CL limit with 200 tonne-years, and the 90\% CL exclusion neutrino fog level.}
\end{figure}
In Figure \ref{fig:foglevels}, we compare these various criteria for the 
neutrino fog, made with the help of the code provided publicly by C.A.J.~O'Hare at \cite{NeutrinoFoggithub},
and the available published projections for the proposed experiments XLZD with 200 and 1000 tonne-years of exposure, and PandaX-xT with 200 tonne-years.
It is useful to note that the 90\% CL exclusion reach cross-sections for the proposed experiments can be a factor of a few below the discovery fog definitions. 
The current LZ2024 exclusions are less than a factor of 4 away from the 5$\sigma$ discovery neutrino fog level, but are still more than an order of magnitude above the 
90\% CL exclusion fog level.

We will use these criteria below to estimate what gaugino parameters can be expected to remain after future dark matter direct detection experiments, with implications for collider discovery aspects discussed in the next section.

\subsection{Models with gaugino mass unification}

The apparent unification of gauge couplings in the MSSM near $2\times 10^{16}$ GeV is a strong motivation to consider models in which the running gaugino masses $M_1$, $M_2$, and $M_3$ also unify at that scale. In this predictive framework, the neutral higgsino LSP mixes predominantly with the bino, and the magnitude of this mixing is constrained by the LZ2024 direct dark matter search bounds. As noted in ref.~\cite{Martin:2024ytt}, the bound becomes monotonically stronger for increasing sign$(\mu)/\tan\beta$, so that the weakest bounds are obtained for small $\tan\beta$ and $\mu<0$, the strongest bounds are obtained for small $\tan\beta$ and $\mu>0$, and large $\tan\beta$ gives an intermediate bound. With the well-motivated assumption of gaugino mass unification, these bounds translate into lower bounds on the bino and gluino masses.

\begin{figure}[t]
\mbox{
  \includegraphics[width=0.515\linewidth]{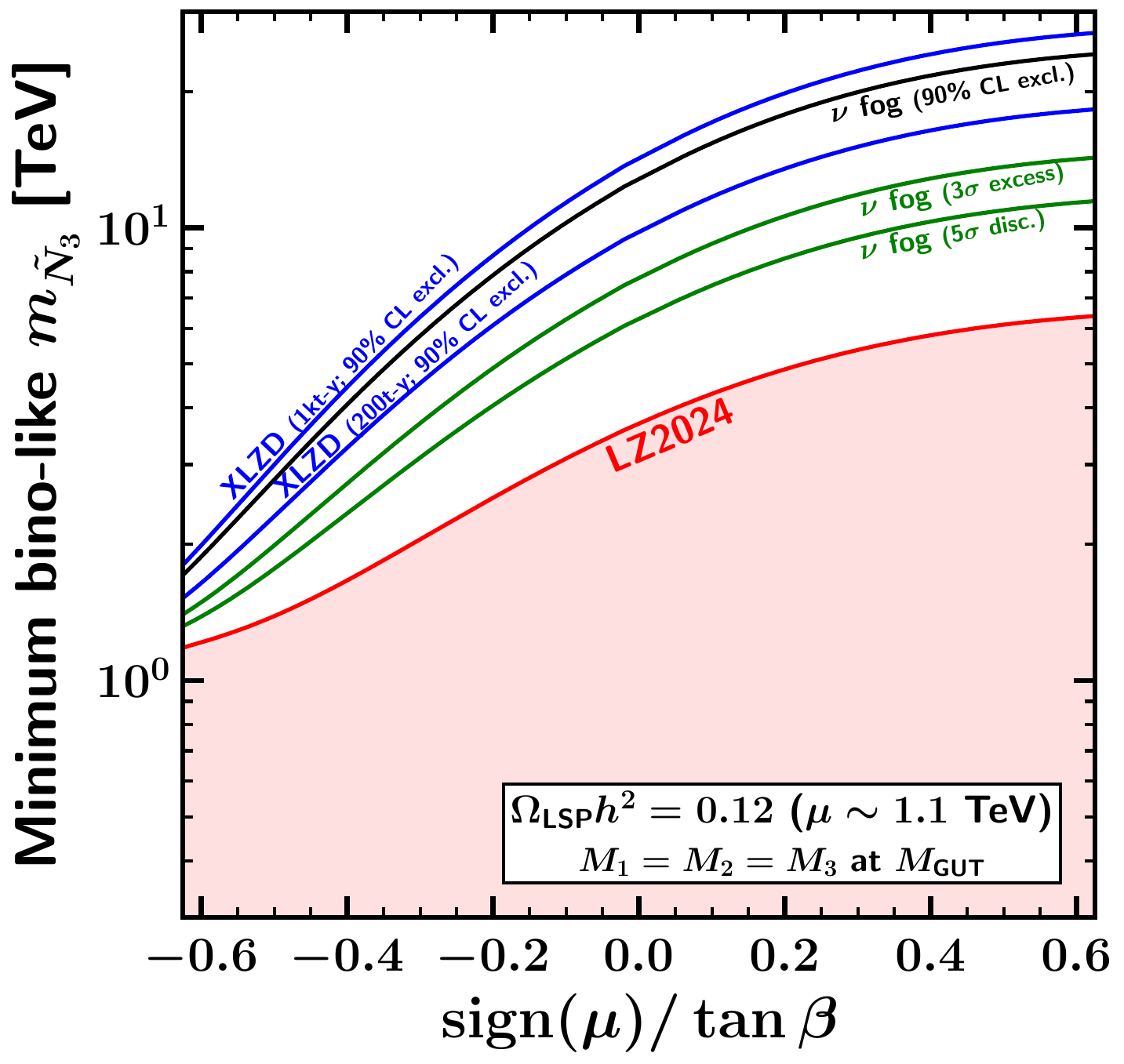}
  \includegraphics[width=0.515\linewidth]{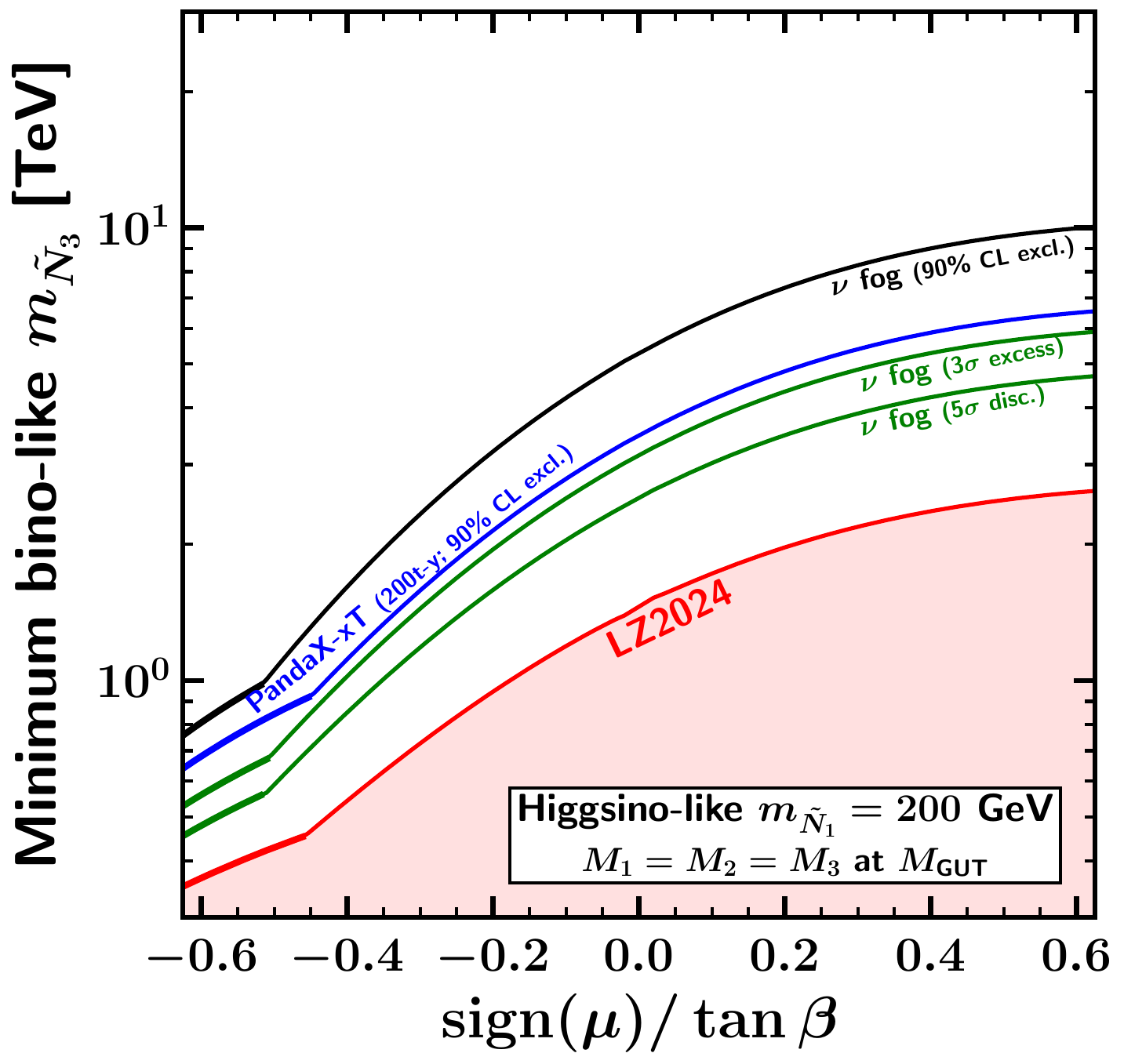}
}
\mbox{
  \includegraphics[width=0.515\linewidth]{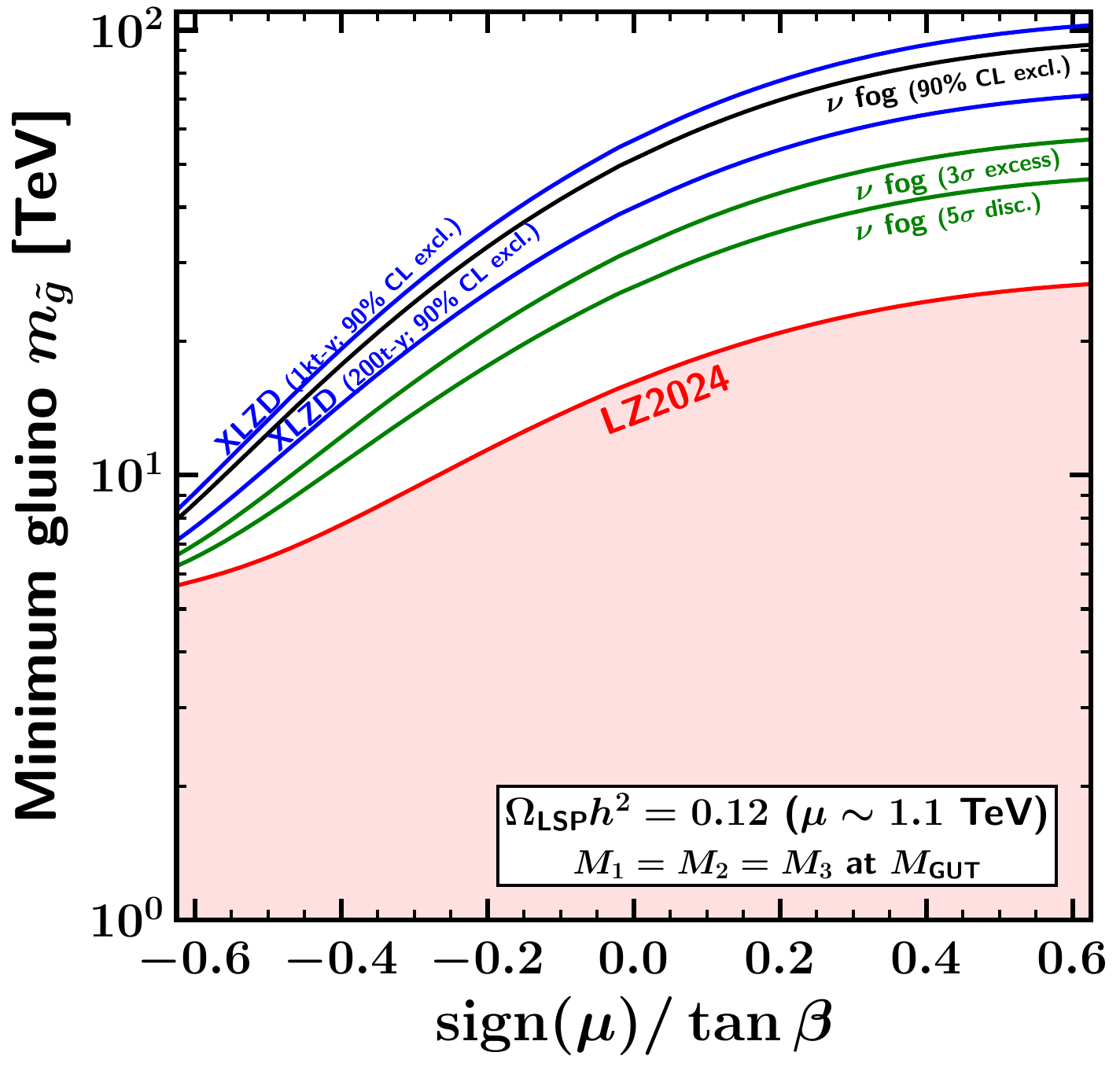}
  \includegraphics[width=0.515\linewidth]{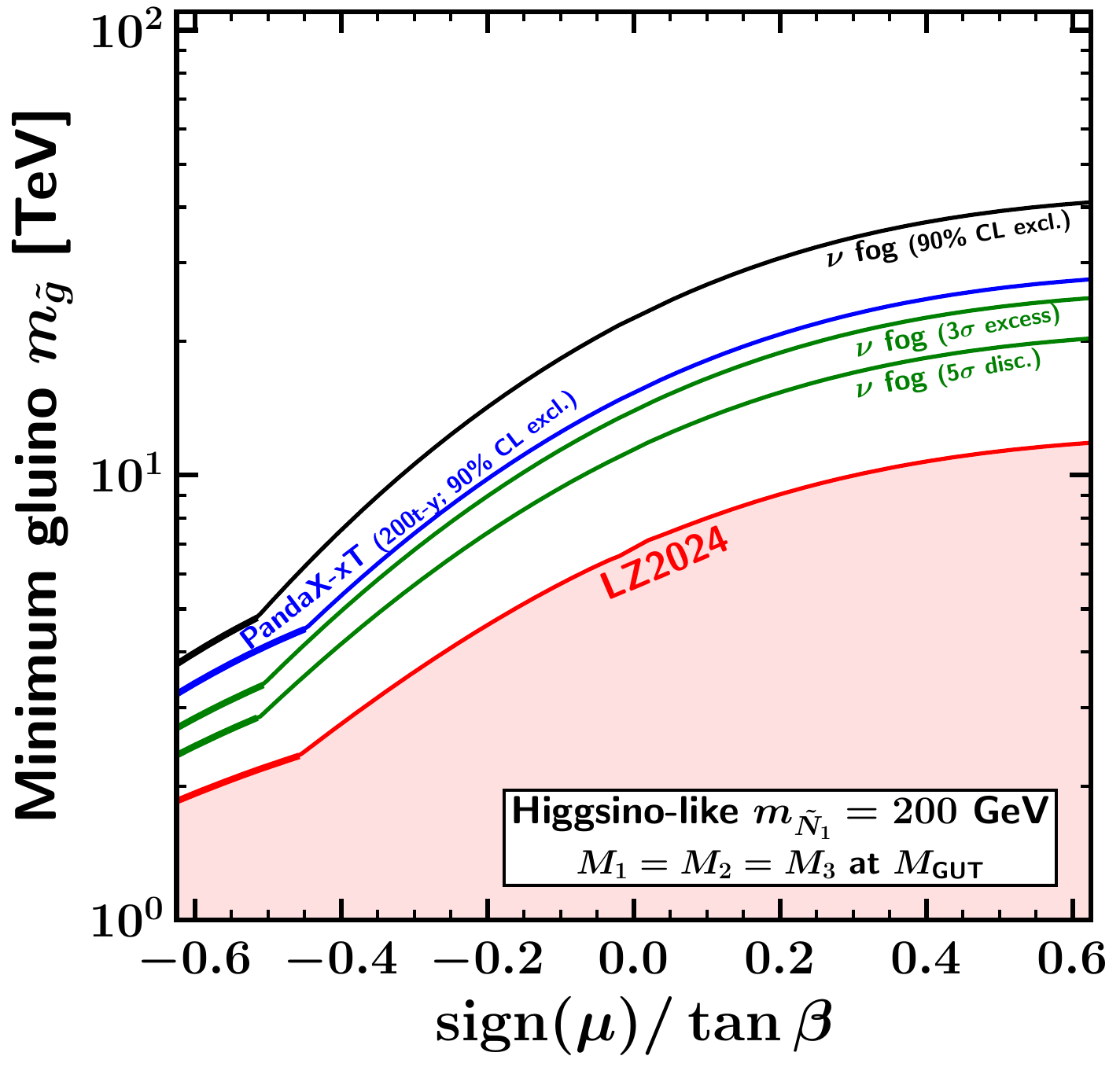}
}
\caption{\label{fig:M1M3minCMSSM} Minimum values of the bino mass $m_{\tilde N_3} \approx M_1$ (top) and the gluino pole mass $m_{\tilde g}$ (bottom) allowed by dark matter direction detection bounds in models where gaugino masses are assumed to unify at $M_{\rm GUT} = 2\times 10^{16}$ GeV.
The left panels assume higgsino-like LSP with mass near 1.1 TeV so that $\Omega_\text{LSP} h^2 = 0.12$, while the right panels fix the higgsino-like LSP mass to be $m_{\tilde N_1} = 200$ GeV with $\Omega_\text{LSP} h^2$ fixed by the thermal relic abundance.
All squark and slepton and heavy Higgs boson masses are taken to be equal to the gluino mass. The shaded region in each panel is the current LZ2024 exclusion. The other lines are various neutrino fog definitions and future exclusion projections as labeled.
For the bolder portion of each curve with $\mu<0$ and $\tan\beta \lesssim 2$ in the right panels, the minimum gaugino masses are determined by the SDn cross-section in each case.}
\end{figure}

We show the results for the minimum allowed bino-like neutralino mass and the minimum allowed gluino pole mass in the left panels of Figure \ref{fig:M1M3minCMSSM}, assuming that the higgsino-like LSP has a mass near 1.1 TeV, adjusted so that the higgsino thermal relic abundance is near the observed dark matter density $\Omega_\text{LSP} h^2 = 0.12$.
We note that even in the most optimistic case of $\tan\beta \approx 1.6$ and negative $\mu$, the bino-like state $\tilde N_3$ must have a mass above 1.2 TeV, and at least several TeV for larger sign$(\mu)/\tan\beta$. The gluino mass would have to be above 5.5 TeV at least, and for large $\tan\beta$, and for positive $\mu$, the gluino mass would have to be at least 15 TeV. We also show in these plots how these bounds should evolve in the future as direct dark matter detection experiments encounter the neutrino fog and projected exclusion levels, using the criteria shown in Figure \ref{fig:foglevels} for the spin-independent cross-section. (In the present case of a 1.1 TeV higgsino, the spin-dependent cross-section is always weaker.) The power of future direct detection experiments is such that, given a projected 90\% CL exclusion from XLZD with 1 kt-year of exposure, the gluino mass could be inferred to exceed 100 TeV for small $\tan\beta$ and positive $\mu$, and 55 TeV for large $\tan\beta$, and 8 TeV even for small $\tan\beta$ and negative $\mu$. Conversely, and more optimistically, a hint or detection of dark matter direct detection would give some indirect indication of the possible range of gaugino masses responsible for the higgsino mixing leading to its coupling to nucleons.

If the higgsino is lighter than 1.1 TeV, its predicted thermal abundance is such that it would make up only a part of the dark matter called for by astrophysical observations. This means that the direct detection cross-section limits are weaker by a factor of $\Omega_\text{LSP} h^2/0.12$. Nevertheless, the LZ2024 limits still constrain the purity of the higgsino LSP, and therefore the gaugino masses, in a quite nontrivial way. In the right panels of Figure \ref{fig:M1M3minCMSSM}, we show the current bounds, and the neutrino fog and exclusion projections, assuming as an example that the higgsino-like LSP mass is 200 GeV (chosen because below this is where current collider bounds begin to come into play). These bounds are weaker than for the critical case of a 1.1 TeV higgsino. We note that eventually when direct detection experiments encounter the neutrino exclusion fog, a higgsino that makes up even part of the dark matter would be forced to be so pure that the
gluino mass would have to be almost 4 TeV, assuming gaugino mass unification.

\clearpage

\subsection{Anomaly mediated supersymmetry breaking models}

\begin{figure}[t] 
\mbox{
  \includegraphics[width=0.515\linewidth]{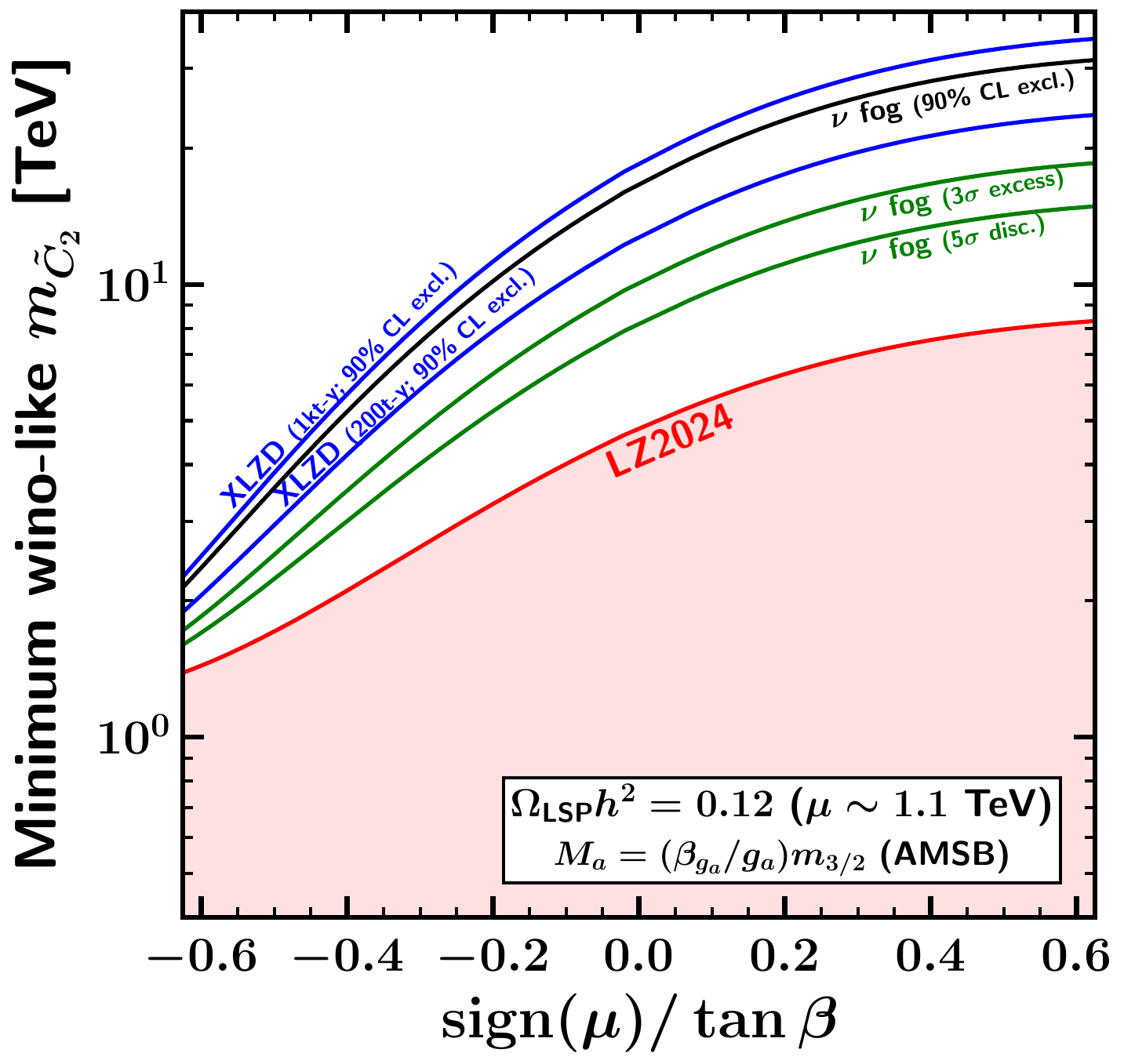}
  \includegraphics[width=0.515\linewidth]{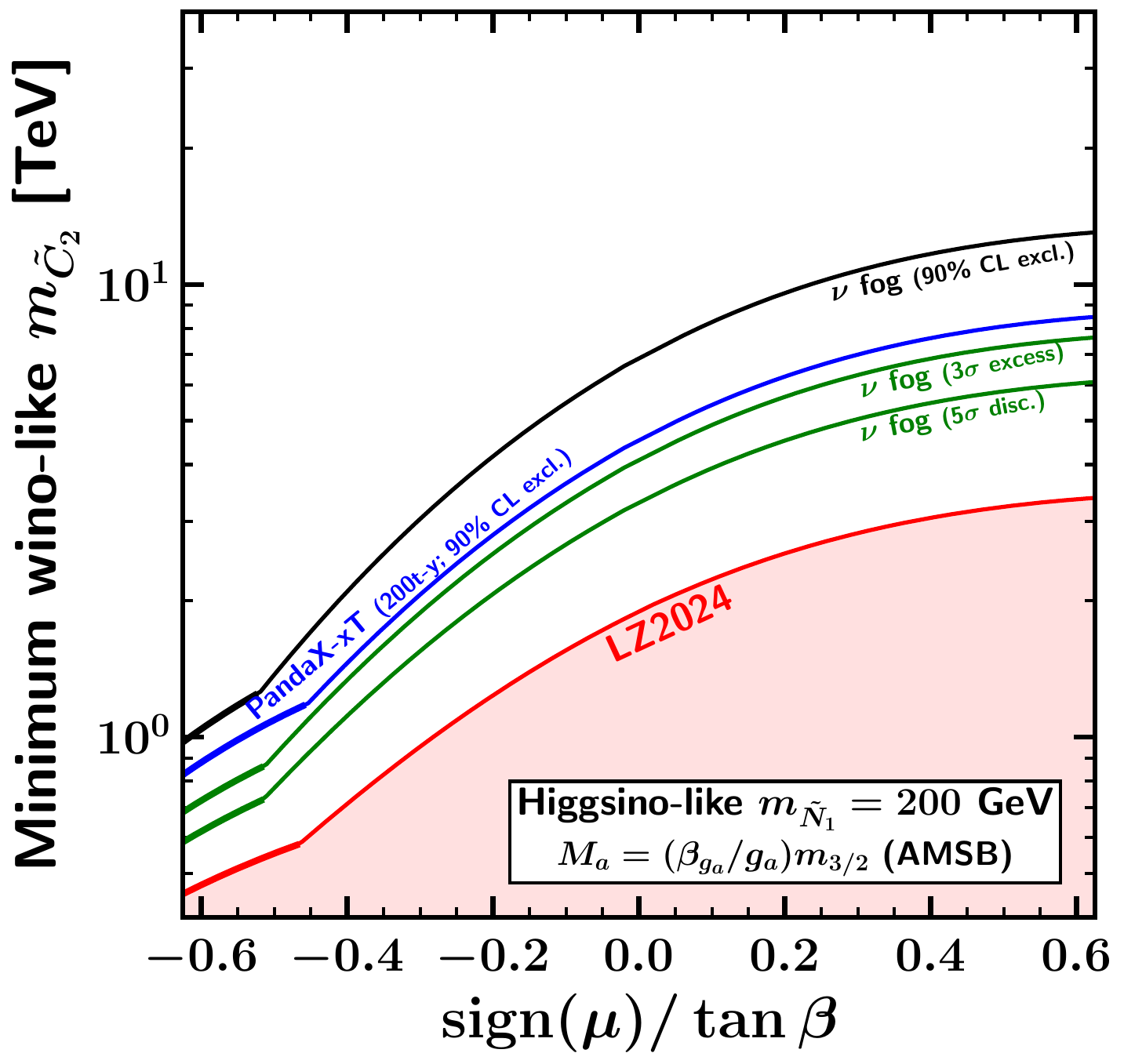}
}
\caption{
\label{fig:M2minAMSB}Minimum wino-like chargino mass $m_{\tilde C_2}$ allowed by
dark matter direction detection bounds, in AMSB models. The left panel assumes higgsino-like LSP with mass near 1.1 TeV so that $\Omega_\text{LSP} h^2 = 0.12$, while the right panel fixes the higgsino-like LSP mass to be $m_{\tilde N_1} = 200$ GeV, which turns out to imply $\Omega_\text{LSP} h^2 \approx 0.004$ to $0.007$.
All squark and slepton and heavy Higgs boson masses are taken equal to the gluino mass. In each panel, the shaded region is the result for the current LZ2024 bounds, and the other lines assume various neutrino fog definitions and future exclusion projections as labeled.
For the bolder portion of each curve with $\mu<0$ and $\tan\beta \lesssim 2$ in the right panel, the minimum wino mass is determined by the SDn cross-section.}
\end{figure}

Another, quite different, motivated scenario for the gaugino mass ratios is provided by anomaly-mediated supersymmetry breaking (AMSB) models. AMSB predicts that the wino is much lighter than the bino  ($M_1 = 3.2 M_1$ near the TeV scale), and is therefore mostly responsible for the mixing and mass splitting of higgsinos. The higgsino purity constraints implied by the LZ2024 dark matter direct detection bounds therefore impose a lower bound on the wino-like chargino and neutralino masses, which are shown in Figure \ref{fig:M2minAMSB} for the cases that $\Omega_\text{LSP} h^2 = 0.12$ (so that the higgsino masses are near 1.1 TeV) and that the higgsino-like LSP mass is fixed at $m_{\tilde N_1} = 200$ GeV (so that the thermal relic density is $\Omega_\text{LSP} h^2 \approx 0.004$ to $0.007$).

Although the bounds are much weaker for lighter higgsino masses due to the dark matter direct detection suppression by $\Omega_\text{LSP} h^2/0.12$, they are still nontrivial even if $m_{\tilde N_1}$ is only 200 GeV, especially if $\mu$ is positive. Besides the LZ2024 bounds, we also show for comparison some future projected limits and different definitions of the neutrino fog.

\section{Detecting higgsinos at colliders\label{sec:HDMatcolliders}}
\setcounter{equation}{0}
\setcounter{figure}{0}
\setcounter{table}{0} 
\setcounter{footnote}{1}

The detection or exclusion of higgsinos at the LHC is well-known to be a difficult challenge, for several reasons. First, the production cross-section is limited due to having no direct tree-level interactions with the gluon or light quark partons in the proton. Second, the mass splittings among higgsinos are quite small, leading to low efficiencies to pass triggers and signal cuts. This problem is particularly acute if the LSP is stable and has a thermal relic abundance, because then the purity constraints following from the LZ direct detection limits require that the mass splittings are strongly bounded from above, as shown in refs.~\cite{Martin:2024pxx,Martin:2024ytt}. Searches for disappearing tracks \cite{ATLAS:2022rme,CMS:2023mny} can be effective if the mass splitting  between the higgsino-like chargino and the neutral LSP, $\Delta M_+ = m_{\tilde C_1} - m_{\tilde N_1}$, is less than about 350 MeV, but this requires very pure higgsinos. Searches for soft leptons with  missing transverse energy \cite{ATLAS:2019lng,ATLAS:2021moa,CMS:2021edw,CMS:2024gyw} have a significant reach for large $\Delta M_+$, but become severely limited for mass splittings consistent with a dark matter interpretation, and disappear completely for $\Delta M_+$ below 1 GeV.  The intermediate range of $\Delta M_+$ from about 300 MeV to 1.2 GeV, which is a prime target based on the constraints from dark matter direction detection, can be targeted with mildly displaced track searches \cite{ATLAS:2024umc,CMSdisplacedsoft}. Very recently, low-momentum lepton tracks from $\tilde N_2 \rightarrow \tilde N_1 Z^*$ have been used \cite{CMS:2025mie,ATLAS:2025lhc} to probe the range of $\Delta M_0$ from a few GeV down to about 1 GeV, but with a limited mass range $m_{\tilde N_2} < 130$ GeV. Another search strategy is to look for monojets with missing energy, as in refs.~\cite{ATLAS:2021kxv,CMS:2021far}.

However, all LHC limits on directly produced higgsinos with a stable LSP are currently limited to masses below about 200 GeV. For larger mass splittings of order 10-30 GeV, there are slight excesses in the soft lepton channels \cite{ATLAS:2019lng,ATLAS:2021moa,CMS:2021edw,CMS:2024gyw}, manifesting as weaker-than-expected limits, but these mass splittings are too large to be compatible with the higgsino dark matter interpretation in the MSSM. They could be accommodated in a variety of ways if the LSP is unstable, or mixed with a light singlino fermion, or
is a wino-bino mixture. Some recent papers exploring various possible supersymmetric model bases for these mild excesses can be found in refs.~\cite{Agin:2023yoq,Agin:2024yfs,Agin:2025vgn,Araz:2025bww}, which noted potentially overlapping  excesses in monojet searches, and refs.~\cite{Chakraborti:2024pdn}, \cite{Martin:2024pxx,Martin:2024ytt}, and \cite{Ellwanger:2024vvs}. Prospects for higgsinos and far-future colliders are discussed in refs.~\cite{Canepa:2020ntc,Capdevilla:2024bwt}.

\begin{figure}[t]
  \includegraphics[width=0.49\linewidth]{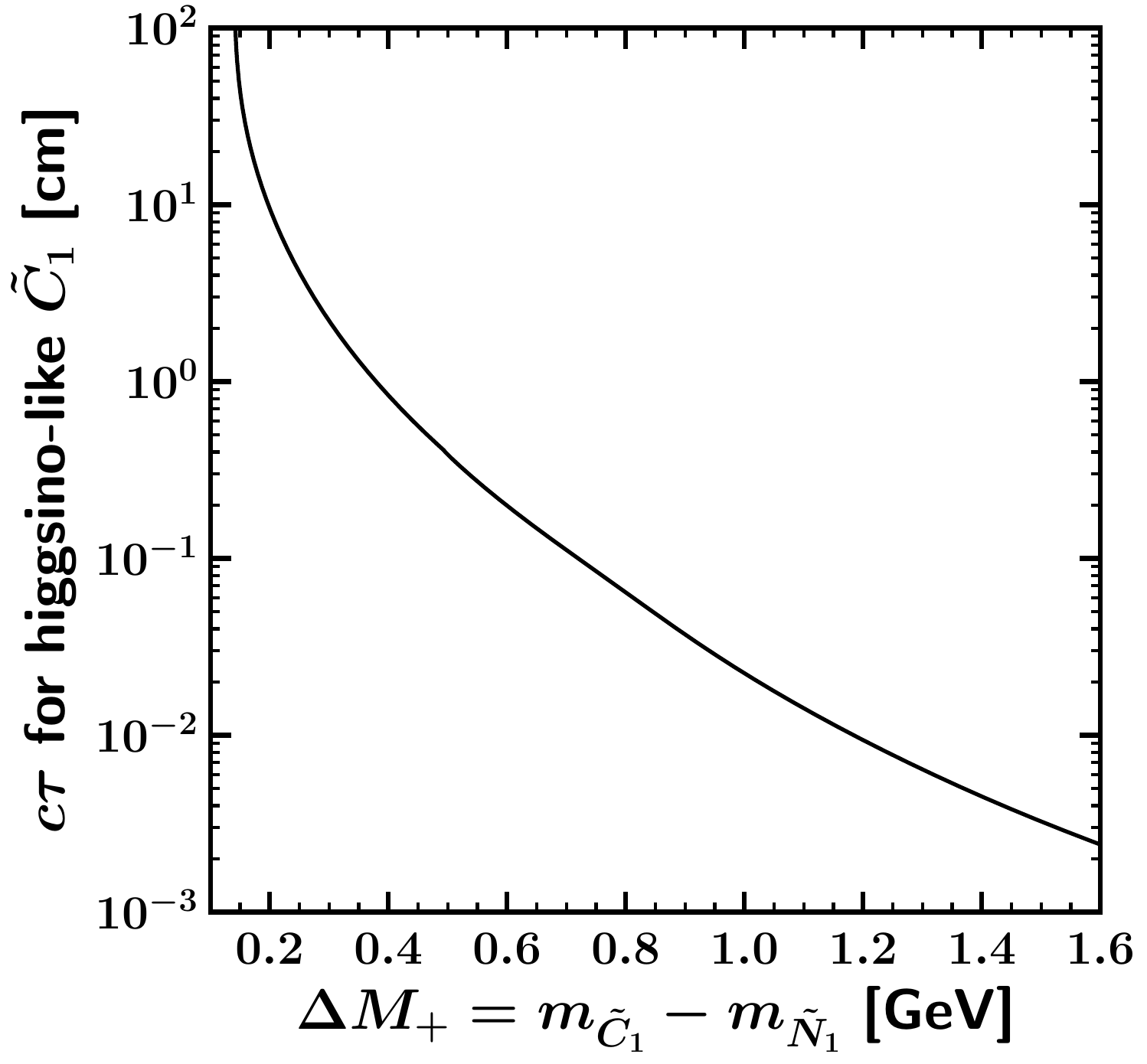}
  \includegraphics[width=0.49\linewidth]{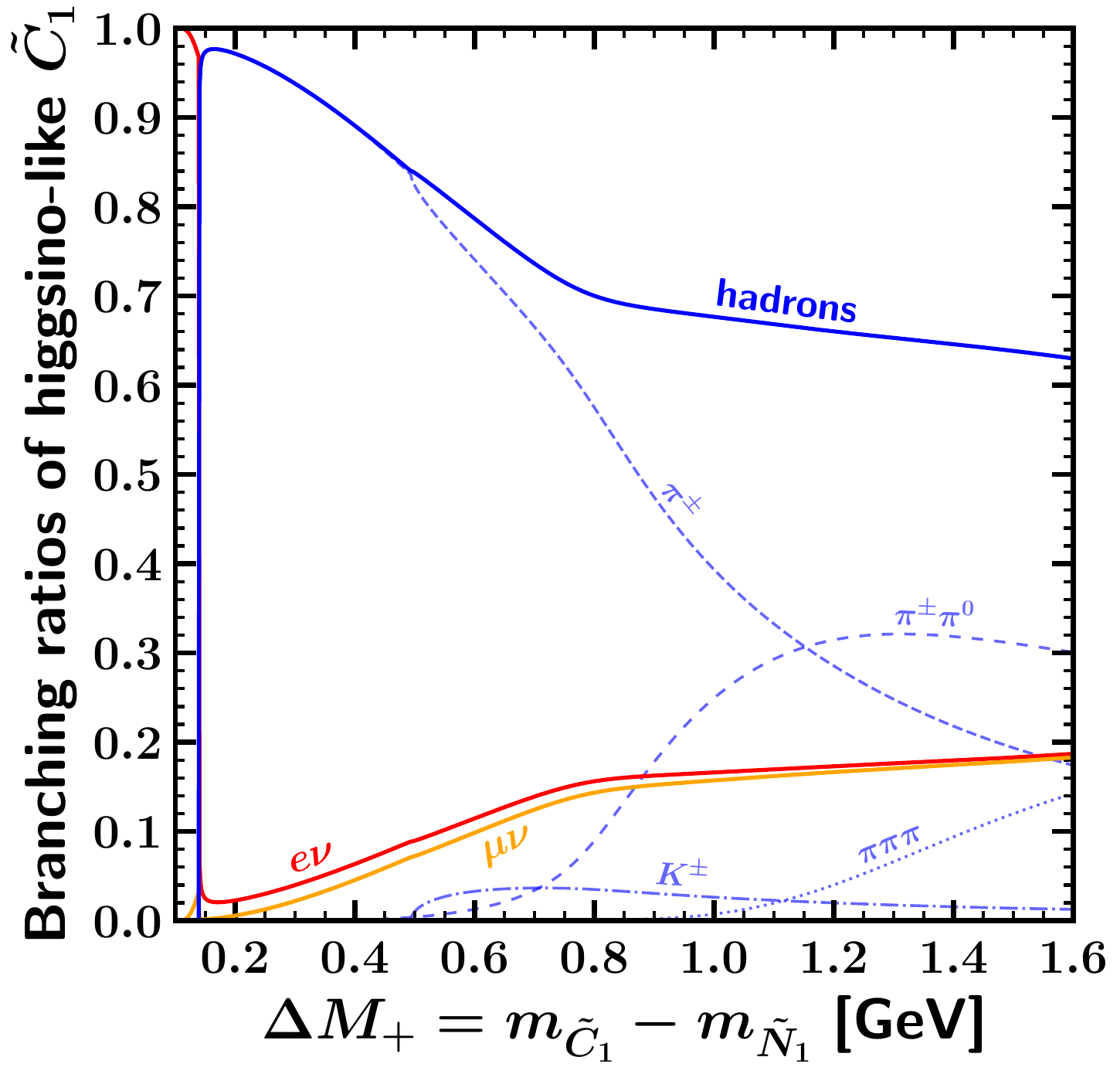}
\caption{\label{fig:ctau_BR} The decay length $c\tau$ in cm (left panel) and branching ratios (right panel) for charged higgsinos as a function of the
mass difference $\Delta M_+ = m_{\tilde C_1} - m_{\tilde N_1}$.}
\end{figure}

The decay lifetime of a charged higgsino-like state to its neutral LSP partner is a crucial parameter for searches, and depends directly on the mass difference $\Delta M_+$. In Figure \ref{fig:ctau_BR}, we give the proper decay distance $c\tau$, and the branching ratios into various final states, as a function of $\Delta M_+$, using results obtained in refs.~\cite{Thomas:1998wy,Chen:1996ap,Ibe:2023dcu}, including final states with multiple mesons. 
(The first of these references obtained results for a completely unmixed neutral state; the contributions to $c\tau$ for higgsinos in the present situation are multiplied by 2 to account for the fact that one of the neutral higgsinos is heavier than the charged state,
and thus decouples from its decay.)
This shows why disappearing track searches are effective for mass splittings near 300 MeV,
while displaced track searches are viable for larger mass splittings, but become ineffective
for mass splittings somewhat larger than 1 GeV.

One of the main difficulties for such searches is the low cross-section for signal regions after triggers and cuts are taken into account, which perhaps can be avoided if the higgsinos are produced from the decays of other superpartners. However, in the full or partial dark matter interpretation, this is where the purity constraints
from direct dark matter detection come into play, requiring gauginos that could be produced with larger cross-sections to be heavy, as discussed in the previous section.

\subsection{Models with gaugino mass unification}

In the models with the well-motivated assumption of gaugino mass unification, where bino is the lightest electroweakino, we first consider the prospects for gluino pair-production as a means to probe higgsino dark matter. However, as seen in Figure~\ref{fig:M1M3minCMSSM}, in the case where higgsino LSP makes up for all of the dark matter, the gluino mass would have to be at least 5.5 TeV even in the most optimistic case of negative $\mu$ and $\tan \beta \approx 1.6$, dashing any hopes of producing it at the LHC with the current energy configuration. This only becomes more restrictive for larger sign$(\mu)/\tan\beta$, and/or as direct detection limits approach the neutrino fog or future projections.

On the other hand, if the higgsino is lighter than 1.1 TeV, the direct detection bounds are weaker, and this leaves room for gluino pair production to be a factor at the LHC, but only if $\tan \beta$ is small and $\mu$ is negative. This can be inferred, for example, from the bottom-right panel of Figure~\ref{fig:M1M3minCMSSM}, where the higgsino LSP mass is fixed at 200 GeV. Here, it is useful to note that existing searches for gluino pair production do not focus on the higgsino LSP scenario. Based on our findings, it would be useful for the LHC detector collaborations to consider especially targeting $pp \rightarrow \tilde g \tilde g$ searches with subsequent cascade decays through gauginos to the higgsinos, using some benchmark model lines with, say, $\tan\beta=2$, $\mu<0$, and varying $|\mu|$ roughly in the vicinity of 200 GeV, in order to fill in this possibility that remains from the current direct detection limits. Finally, as the direct detection bounds encounter the neutrino fog, the gluino would again be out of reach of the LHC without an energy upgrade.

Due to the effects of renormalization group running, in most cases the squark masses cannot be much lighter than the gluino, and so the pessimistic projections about accessibility at the LHC applies to them as well. A notable possible exception is the lighter top squark, because radiative effects from the top-quark Yukawa coupling and $\tilde t_R/\tilde t_L$ mixing both tend to reduce one of the top-squark mass eigenvalues. If this effect is great enough, one can envision the lighter state $\tilde t_1$ as a good target for LHC searches.

\begin{figure}
\begin{minipage}[]{0.6\linewidth}
\begin{flushleft}
  \includegraphics[width=0.98\linewidth]{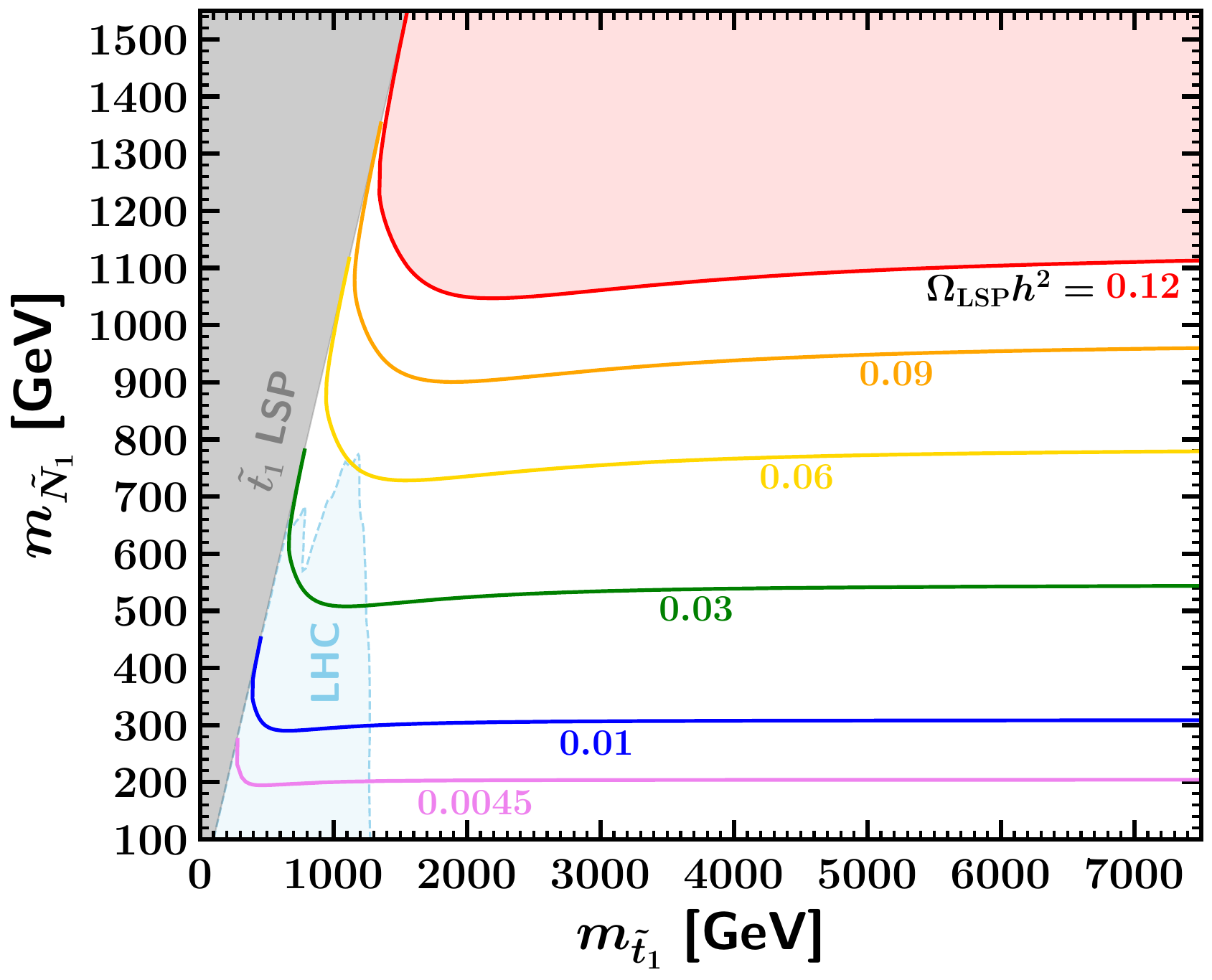}
  \end{flushleft}
\end{minipage}
\begin{minipage}[]{0.39\linewidth}
\caption{\label{fig:stop1N1plane} Higgsino-like LSP mass $m_{\tilde{N}_1}$ as a function of the stop mass $m_{\tilde{t}_1}$ that reproduces various values of $\Omega_\text{LSP} h^2$, as labeled, for $\tan \beta = 1.6$ and $\mu < 0$. The gaugino mass parameters, and all other scalar masses (except $M_h$), are set to 10 TeV. In order to show a rough estimate of the Run 2 LHC sensitivity with 139 fb$^{-1}$ at $\sqrt{s}=13.6$ TeV in the absence of reported searches targeting this specific scenario, we show as the light blue shaded region
the reported exclusions for sbottoms from ref.~\cite{ATLAS:2021yij}.
} 
\end{minipage}
\end{figure}

If $\tilde t_1$ is light enough to be accessible to the LHC, one might also worry about its effect on the predicted $\Omega_\text{LSP} h^2$ due to co-annihilation effects in the early universe. In fact, the stop-coannihilation region for dark matter is continuously connected in parameter space to the higgsino-coannihilation region that sets $\Omega_\text{LSP} h^2 = 0.12$ for $m_{\tilde N_1} \sim 1.1$ TeV. This is illustrated in Figure \ref{fig:stop1N1plane}, which shows contours of equal values of the thermal $\Omega_\text{LSP} h^2$ from 0.0045 (where the higgsino would make up less than 4\% of the dark matter) up to 0.12 (where the higgsino would make up all of the dark matter). 
In making this plot, we have chosen $\mu<0$ and $\tan\beta = 1.6$, so that the higgsino purity constraints from LZ2024 are weakest, but the results are not strongly dependent on this choice. We find that because of a negative interference effect in the coannihilation channel $\tilde N_2 \tilde C_1^+ \rightarrow t \overline b$ with a virtual $t$-channel $\tilde t_1$, the critical value needed to obtain $\Omega_\text{LSP} h^2 = 0.12$ is reduced slightly for moderate $m_{\tilde t_1}$, of order 2 TeV. However, for even lower stop masses the usual stop coannihilation mechanism becomes important, and the critical higgsino mass is raised substantially, giving a hook shape to the lines in the figure. This requires that $m_{\tilde t_1} - m_{\tilde N_1}$ is small.

\begin{figure}
\begin{minipage}[]{0.6\linewidth}
\begin{flushleft}
  \includegraphics[width=0.95\linewidth]{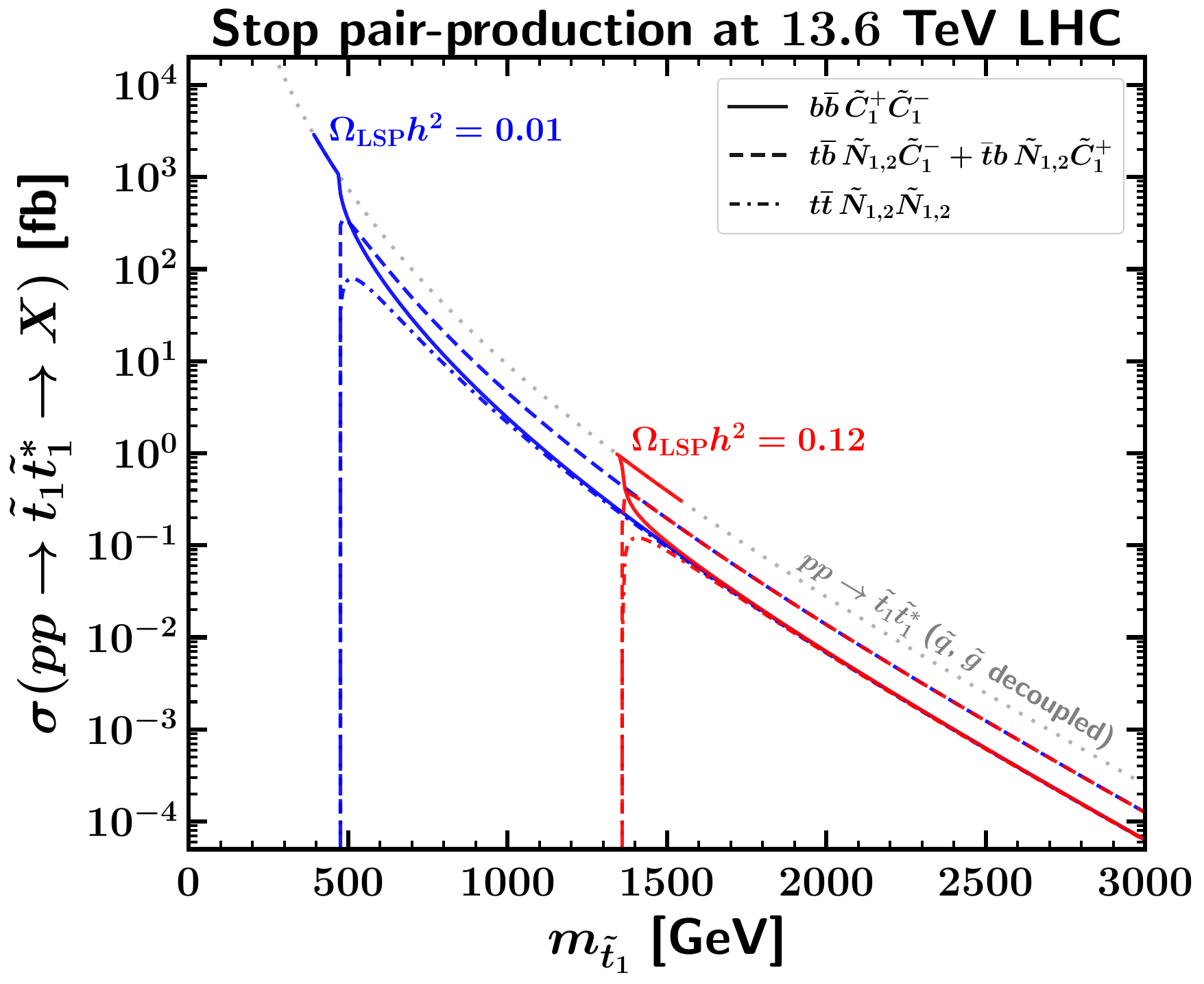}
\end{flushleft}
\end{minipage}
\begin{minipage}[]{0.39\linewidth}
\caption{\label{fig:stopsigmaBR} Cross-section times branching ratio ($\sigma \times \text{BR}$) for the final states in eq.~(\ref{eq:stopsigmabR}) from stop pair-production at 13.6 TeV LHC, as a function of the stop mass $m_{\tilde{t}_1}$ that reproduces $\Omega_\text{LSP} h^2 = $ 0.12 (red) and 0.01 (blue), for $\tan \beta = 1.6$ and $\mu < 0$. The gaugino mass parameters, and all other scalar masses (except $M_h$), are set to 10 TeV. In a small portion of the $\Omega_\text{LSP} h^2 = 0.12$ case near $m_{\tilde t_1}= 1500$ GeV, there are two lines for a given stop
mass; the upper segment is the stop-coannihilation case.}
\end{minipage}
\end{figure}

Following direct pair production of top squarks, the flavor-unsuppressed decays to higgsinos are
\beq
\tilde t_1 &\rightarrow& 
\left \{ \begin{array}{ll}
b \tilde C_1^+,\qquad\! (\geq 50\%),
\\
t \tilde N_2,\qquad (\leq 25\%),
\\
t \tilde N_1,\qquad (\leq 25\%),
\end{array}
\right .
\eeq
in which the $\geq$ and $\leq$ symbols become equalities in the limit that there is no kinematic suppression of the decays to top quarks, so $m_{\tilde t_1} - m_{\tilde N_2} \gg m_t$. In the opposite extreme, if $m_{\tilde t_1} - m_{\tilde N_2} < m_t$, then the decays to bottom quarks dominate. The LHC processes are therefore
\beq
pp &\rightarrow& \tilde t_1 \tilde t_1^*  \>\rightarrow\> 
\left \{ \begin{array}{ll}
b \overline b \tilde C_1^+ \tilde C_1^-,
\\
t \overline b \tilde N_{1,2} \tilde C_1^-,
\\
\overline t b \tilde N_{1,2} \tilde C_1^+,
\\
t \overline t \tilde N_{1,2} \tilde N_{1,2},
\end{array}
\right .
\label{eq:stopsigmabR}
\eeq
followed by $\tilde C_1$ and $\tilde N_2$ decaying to the LSP with soft additional tracks that may or may not be discernable. The cross-section times branching ratio results are shown in Figure \ref{fig:stopsigmaBR}, along with the total cross-section as the dotted line. 

Present LHC top-squark searches do not seem to be dedicated to the precise scenario, as it is generally assumed instead that the charginos decay to $W\tilde N_1$. The exception is a somewhat older search in ref.~\cite{ATLAS:2017avc}, based on 36 fb$^{-1}$ at $\sqrt{s} = 13$ TeV, which includes a search in a simplified model but in which the $\Delta M_+$ is fixed at 1 GeV. If the soft chargino decay products are not observed, then the $b \overline b \tilde C_1^+ \tilde C_1^-$ signal in eq.~(\ref{eq:stopsigmabR}) is similar to the simplified model searches for bottom squarks decaying to binos, for which limits have been reported in refs.~\cite{ATLAS:2017avc,ATLAS:2021yij,CMS:2019zmd}. The other final states resemble the more traditional searches for top squarks. A dedicated study of the signal efficiencies is beyond the scope of this paper, and will presumably become obsolete on a short time scale as experimental LHC search results become available, but we suspect that the reach for top squarks decaying to higgsinos should be at least roughly similar to the existing ones  for bottom squarks and top squarks decaying to winos and binos. To give some indication of the region which might be accessible with LHC data so far, but in the absence of dedicated searches for these models, we have therefore shown as a shaded region in Figure \ref{fig:stop1N1plane} the exclusion region for sbottoms from ref.~\cite{ATLAS:2021yij}. For the longer term, we note that the theoretical study in ref.~\cite{Baer:2023uwo} indicates that  discovery prospects for top-squark pair production at a high-luminosity LHC with 3000 fb$^{-1}$ at $\sqrt{s} = 14$ TeV may extend up to $m_{\tilde t_1} = 1.7$ TeV, with exclusion possible up to 2 TeV.

Furthermore, exploiting the possibility of observing additional soft and mildly displaced tracks and disappearing tracks from the $\tilde C_1$ and $\tilde N_2$ decays could enhance the efficiencies for these signal regions. Although the total cross-sections might be smaller, these channels might be more efficient than for direct higgsino pair production because of the presence of jets and leptons from the $b$ or $t$ decays. Estimates of the reach in this case should be undertaken by the experimental collaborations with detailed knowledge of the detectors. We suggest that in the future searches for top-squark pairs decaying directly to higgsino LSPs should be prioritized, and emphasize that they should be generalized to include a larger variety of higgsino mass splittings. We also emphasize again the special relevance of benchmarks with negative $\mu$ and small $\tan\beta$, in view of the dark matter constraints we have discussed above.

\subsection{Anomaly mediated supersymmetry breaking models}

In AMSB models, as before, we can infer from Figure~\ref{fig:M2minAMSB} that the only hope for discovery of gauginos at the LHC occurs for $\mu<0$ and small $\tan\beta$.

\begin{figure}[t]
\begin{minipage}[]{0.6\linewidth}
\begin{flushleft}
  \includegraphics[width=0.96\linewidth]{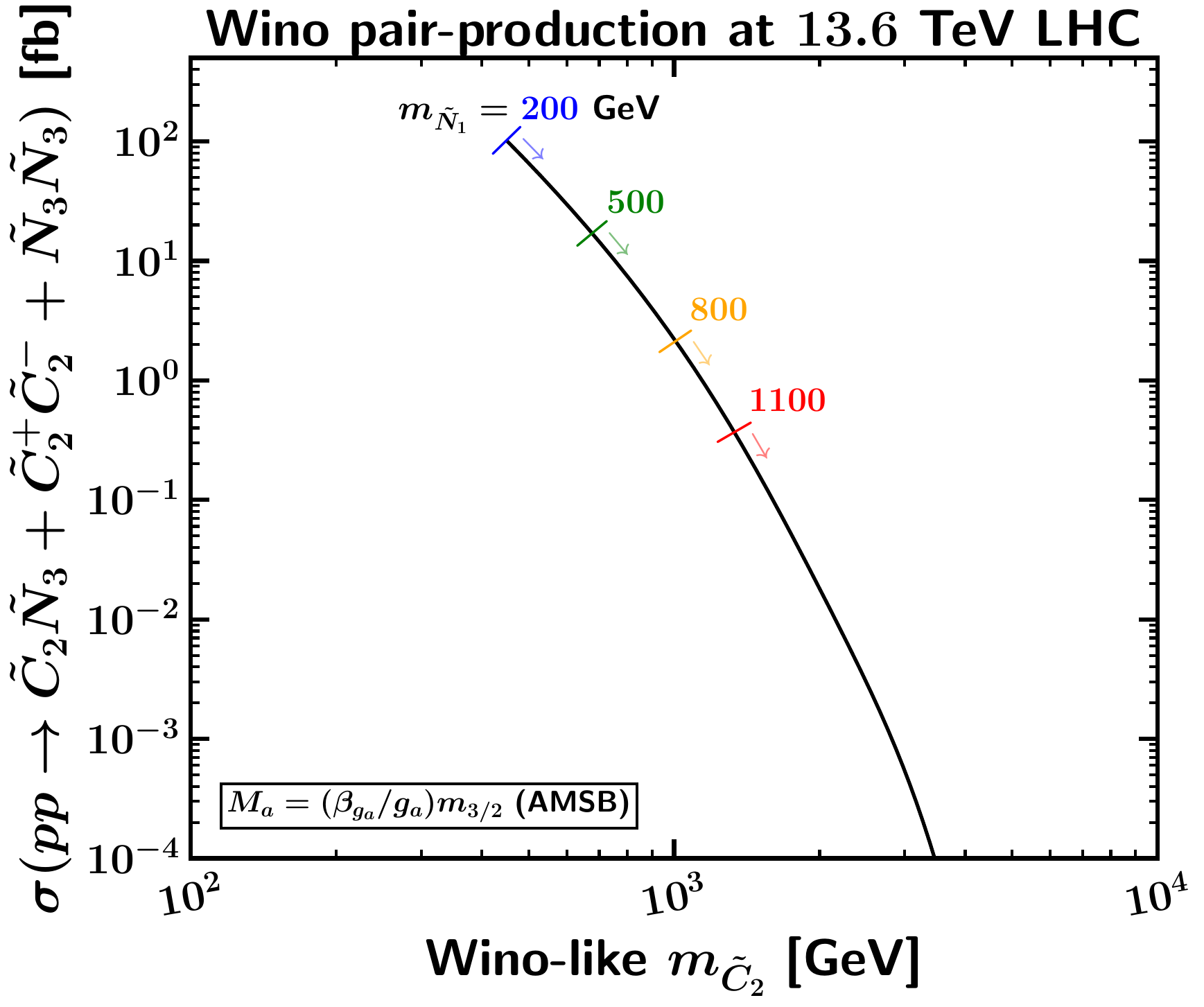}
\end{flushleft}
\end{minipage}
\begin{minipage}[]{0.39\linewidth}
\caption{\label{fig:winocrosssection} The cross-section for wino pair-production, as a function of the wino-like chargino $\tilde C_2$ mass, in AMSB models, assuming
that squarks are decoupled. The portion of the curve allowed by the LZ2024 dark matter direct detection
bounds for various higgsino-like LSP masses with $\text{sign}(\mu)/\tan \beta \geq -0.625$
are indicated by the hash marks and arrows.
}
\end{minipage}
\end{figure}

Besides the ongoing standard searches for direct higgsino pair production, it is interesting to consider the pair production of winos $\tilde W^0 \approx \tilde N_3$ and $\tilde W^\pm \approx \tilde C_2^\pm$, pinning hopes on the parameter space remaining with $\mu<0$ and small $\tan\beta$. The LHC production cross-section with $\sqrt{s} = 13.6$ TeV is shown in Figure \ref{fig:winocrosssection}.
The part of the curve allowed by the LZ2024 dark matter direct detection bounds for various higgsino-like LSP masses with $\mu<0$ and $\tan\beta=1.6$ are indicated by the hash marks and arrows. For larger $\tan\beta$, the allowed wino mass for a given higgsino mass is required to be larger, giving a smaller cross-section. The branching ratios of winos to nearly pure higgsinos are close to
\beq
\tilde C_2 \rightarrow 
\left \{ \begin{array}{ll}
W \tilde N_{1,2},\qquad\!\!\!\!\!\!(50\%)
,
\\
Z \tilde C_1,\qquad(25\%)
,
\\
h \tilde C_1,\,\qquad(25\%)
,
\end{array}
\right .
\qquad
{\rm and}
\qquad
\tilde N_3 \rightarrow 
\left \{ \begin{array}{ll}
W^\pm \tilde C_1^\mp,\!\!\!\!\qquad(50\%)
,
\\
Z \tilde N_{1,2},\qquad(25\%)
,
\\
h \tilde N_{1,2},\,\qquad(25\%)
.
\end{array}
\right .
\eeq
The usual signals from $Wh + \missET$ and $WZ + \missET$ and $W^+W^- + \missET$ familiar from wino-bino searches occur, but with additional soft decay products from the heavier higgsino states. Also, with light higgsinos there is a possibility of same-sign lepton signals from decays through same-sign $W$ bosons, as noted in ref.~\cite{Baer:2013yha}. There are also di-Higgs and $hZ + \missET$ final states, which also do not occur in the traditional wino-bino searches. Wino-pair production with decays to higgsinos have also been considered in refs.~\cite{Baer:2023olq,Baer:2023ech}. The study in ref.~\cite{Baer:2023olq} found that it may be possible to discover winos up to 1.1 TeV, or exclude them up to 1.4 TeV, with the high-luminosity LHC using 3000 fb$^{-1}$ at $\sqrt{s} = 14$ TeV, by combining many different search regions. Here, we note that the new and future dark matter constraints provide particular motivation to conduct these searches with benchmark or simplified models with $\mu<0$ and $\tan\beta \lsim 2$.

\begin{figure}[b]
  \includegraphics[width=13cm]{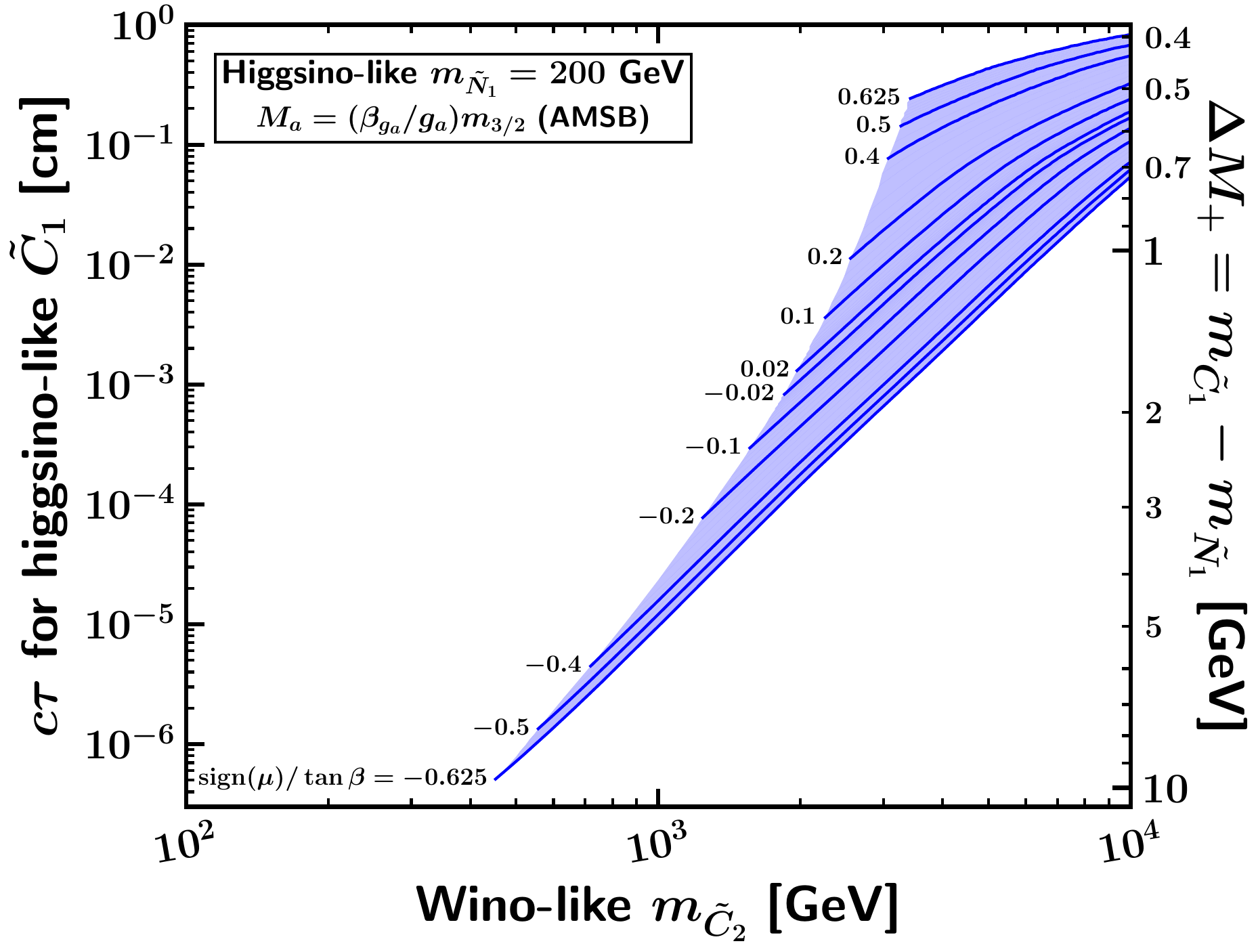}
  \includegraphics[width=13cm]{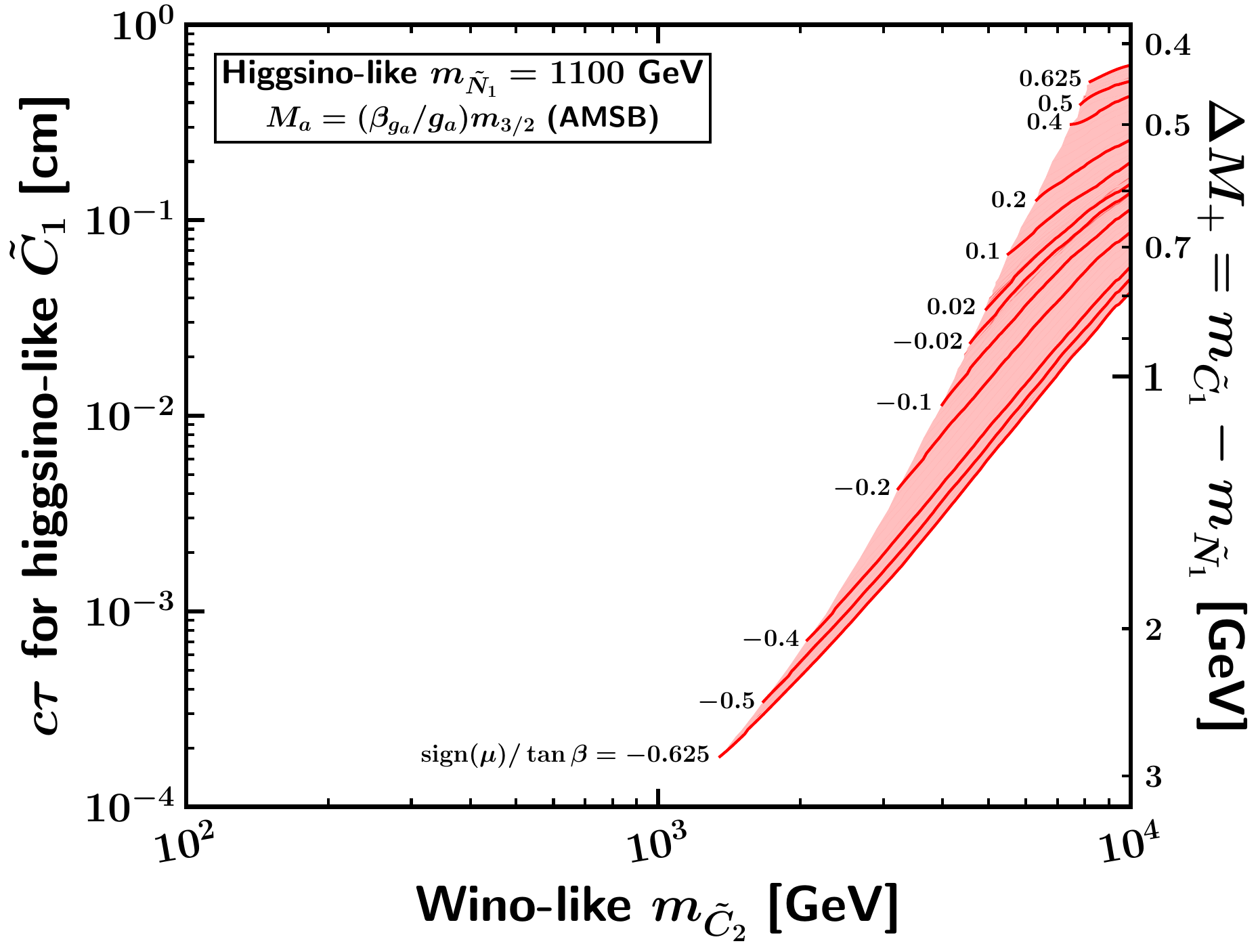}
\caption{\label{fig:ctauAMSB}The proper decay lengths $c\tau$ for a higgsino-like chargino $\tilde C_1$ allowed by the LZ2024 dark matter direct detection bounds, as a function of the wino-like chargino $\tilde C_2$ mass in AMSB models. The different lines correspond to $-0.625 \leq \text{sign}(\mu)/\tan \beta \leq 0.625$ as labeled. The vertical axis on the right side of each panel shows the mass difference $\Delta M_+ = m_{\tilde C_1} - m_{\tilde N_1}$ corresponding to the $c\tau$ on the left vertical axis (as from Figure \ref{fig:ctau_BR}). The top panel has higgsino LSP mass $m_{\tilde{N}_1} = 200$ GeV and the bottom panel has $m_{\tilde{N}_1} = 1100$ GeV.
}
\end{figure}

It is also interesting to consider to what extent exclusion and discovery prospects for winos decaying to higgsinos could be aided by the winos decaying through quasi-stable higgsino-like charginos $\tilde C_1$. In principle, searches for soft and mildly displaced lepton tracks and disappearing tracks can be performed while relying on the energetic leptons and jets from the decays of the Standard Model bosons $W,Z,h$ to provide triggering and efficiencies after cuts to reduce backgrounds. To see to what extent the finite decay length of the higgsino-like $\tilde C_1$ might be leveraged into a signal, we show $c\tau$ in Figure \ref{fig:ctauAMSB} as a function of the wino-like mass $m_{\tilde C_2}$, for AMSB models with higgsino LSP masses $m_{\tilde{N}_1} = 200$ GeV and 1100 GeV, in the regions allowed by LZ2024, showing lines for various $\text{sign}(\mu)/\tan \beta$ between $-$0.625 and 0.625.
Unfortunately, we see that not only do the LZ2024 results bound $m_{\tilde C_2}$ from below, but there is a striking anticorrelation between $m_{\tilde C_2}$ and $c\tau$ evident in Figure \ref{fig:ctauAMSB}, so that when winos are light enough to be produced in significant numbers at the LHC, the decay length of $\tilde C_1$ is much too small to be detectable. This can be understood analytically as follows. The tree-level charged higgsino-LSP mass difference is approximately
\beq
\Delta M_+ &=& \frac{1}{2} m_Z^2 \left [ \frac{c_W^2 (1 - n \sin(2\beta))}{M_2 + n \mu}
+ \frac{s_W^2 (1 + n \sin(2\beta))}{M_1 - n \mu} \right ],
\eeq
where $n = \text{sign}(\mu)$ and $c_W$ and $s_W$ are the cosine and sine of the weak mixing angle. Now, in AMSB models with a higgsino LSP, the first term in square brackets dominates, as $\mu$ is relatively small and $c_W^2/M_2 \gg s_W^2/M_1$. Therefore, $\Delta M_+$ is largest precisely when $n \sin(2\beta)$ is smallest, which occurs when $\mu<0$ and $\tan\beta$ is small. Thus, the dark matter detection constraint from LZ2024 implies that the mass splitting is large and so $c\tau$ for charged higgsinos must be very small for winos accessible to the LHC in AMSB. 

In the preceding, we have assumed for simplicity that $M_2$ and $M_1$ have the same sign, as predicted by AMSB, but the same statement applies to all models with relatively light winos even if $M_2$ and $M_1$ have different signs, as long as the heavy bino approximately decouples. This can be seen from the treatment of the general case in eqs.~(2.3)-(2.11) of ref.~\cite{Martin:2024ytt}. To avoid the anticorrelation for light winos, and allow a small $\Delta M_+$ consistent with a small dark matter direct detection cross-section, one must have $|M_1| < |M_2|$ with opposite signs for $M_1$ and $M_2$, as illustrated by the quasi-blind spot illustrated in Figure 2.2 of ref.~\cite{Martin:2024ytt} and discussed in that paper. This cannot occur in AMSB.

However, while the preceding discussion applies to AMSB models with a higgsino dark matter LSP, the LZ2024 bounds do not apply if the LSP is unstable because of $R$-parity violation. Therefore, it is still certainly worthwhile to look for disappearing tracks or soft displaced tracks within the wino to higgsino decay topologies.

\clearpage

\section{Conclusion\label{sec:conclusion}}
\setcounter{equation}{0}
\setcounter{figure}{0}
\setcounter{table}{0} 
\setcounter{footnote}{1}

Despite years of searching with increasingly powerful probes, higgsino dark matter remains a viable dark matter candidate. However, limits from direct detection probes require that the higgsino have high purity. For that reason we have described the dark matter candidate under consideration in this paper as a ``higgsino" despite necessarily having small components of the fermionic superpartners of the neutral gauge bosons. 

The thermal relic abundance computation requires that a near-pure higgsino have mass approximately 1.1 TeV. Its purity level then can be mapped into the more physical parameters of gaugino masses for a given value of ${\rm sign}(\mu)/\tan\beta$. For unified gaugino masses at the GUT scale, and for anomaly-mediated supersymmetry breaking models, it is sufficient to describe this mixing with a single gaugino mass, as we have done in our analysis above. The higgsino purity constraints translate into substantial multi-TeV, and in some cases tens of TeV, mass constraints on the wino and gluino masses.

The current constraints, and the probing capacity of near-term high mass direct detection experiments, are approaching cross-section levels similar to those of the neutrino fog. Once the neutrino fog limit is reached the discovery capability of direct detection experiments will effectively reach its end. Another approach to discover higgsino dark matter is needed.

As the neutrino fog limit is approached, the higgsino purity limits raise the minimum gaugino masses to a level that will be extremely difficult for the LHC to find, even in its high-luminosity phase. Furthermore, as the purity is raised the mass splitting between the charged higgsino and the neutral higgsino dark matter reduces, making the final decay products of charged higgsino to neutral higgsino decay difficult to find. 

We have found that dedicated searches for wino pair production and top squark pair production will increase the search parameter space for dark matter beyond what direct detection alone can do even after the neutrino fog limit is reached. We have presented how the searches for these two production modes differ somewhat from standard searches for winos and top squarks. In particular, we advocate for searches of stop and wino pairs decaying directly to higgsinos, where the soft decays of $\tilde C_1 \rightarrow \tilde N_1+X_{\rm soft}$ provide both an additional challenge and opportunity to discovery. We expect that future colliders, such as the muon collider or Future Circular Collider (FCC), will increase the discovery capacity through both controls of background (muon collider) or greater kinematic reach (FCC). 

Finally, we remark that there are other modes of higgsino dark matter discovery beyond direct detection and discovery at high-energy colliders. The two most promising are indirect detection possibilities, and searches for electric dipole moments (EDMs).  In the case of indirect detection, there is increased unknown model dependence of the galactic halo profile of the dark matter in our galaxy, especially near its core, which can give orders of magnitude spread in the detection prospects. Nevertheless, it remains an interesting prospect. EDMs are also quite promising~\cite{Co:2022jsn}, but suffer a different type of uncertainty, which is the level of CP violation that exists among superpartner interactions. Although EDMs are one of the most interesting and exciting opportunities to discover evidence of the states within the supersymmetric theory we assume in this article, EDMs require CP violation and this unknown creates unknown probing capacity.

\medskip

{\it Acknowledgments:}
This research was supported in part through computational resources and services provided by Advanced Research Computing (ARC), a division of Information and Technology Services (ITS) at the University of Michigan, Ann Arbor. This work is supported in part by the European Union - Next Generation EU through the PRIN2022 Grant n. 202289JEW4, by the National Science Foundation under grant number 2310533, and by the Department of Energy under grant number DE-SC0007859.



\begin{thebibliography}{}
\baselineskip=15.2pt

\bibitem{Planck:2018vyg}
N.~Aghanim \textit{et al.} [Planck],
``Planck 2018 results. VI. Cosmological parameters,''
Astron. Astrophys. \textbf{641}, A6 (2020)
[erratum: Astron. Astrophys. \textbf{652}, C4 (2021)]
[arXiv:1807.06209 [astro-ph.CO]].

\bibitem{Martin:1997ns}
S.~P.~Martin,
``A Supersymmetry primer,''
Adv. Ser. Direct. High Energy Phys. \textbf{18}, 1-98 (1998)
[arXiv:hep-ph/9709356 [hep-ph]].

\bibitem{Dreiner:2023yus}
H.~K.~Dreiner, H.~E.~Haber and S.~P.~Martin,
``From Spinors to Supersymmetry,''
Cambridge University Press, 2023,
ISBN 978-1-139-04974-0

\bibitem{Arkani-Hamed:2004ymt}
N.~Arkani-Hamed and S.~Dimopoulos,
``Supersymmetric unification without low energy supersymmetry and signatures for fine-tuning at the LHC,''
JHEP \textbf{06}, 073 (2005)
[arXiv:hep-th/0405159 [hep-th]].

\bibitem{Giudice:2004tc}
G.~F.~Giudice and A.~Romanino,
``Split supersymmetry,''
Nucl. Phys. B \textbf{699}, 65-89 (2004)
[erratum: Nucl. Phys. B \textbf{706}, 487-487 (2005)]
[arXiv:hep-ph/0406088 [hep-ph]].

\bibitem{Pierce:2004mk}
A.~Pierce,
``Dark matter in the finely tuned minimal supersymmetric standard model,''
Phys. Rev. D \textbf{70}, 075006 (2004)
[arXiv:hep-ph/0406144 [hep-ph]].

\bibitem{Profumo:2004at}
S.~Profumo and C.~E.~Yaguna,
``A Statistical analysis of supersymmetric dark matter in the MSSM after WMAP,''
Phys. Rev. D \textbf{70}, 095004 (2004)
[arXiv:hep-ph/0407036 [hep-ph]].

\bibitem{Cheung:2012qy}
C.~Cheung, L.~J.~Hall, D.~Pinner and J.~T.~Ruderman,
``Prospects and Blind Spots for Neutralino Dark Matter,''
JHEP \textbf{05}, 100 (2013)
[arXiv:1211.4873 [hep-ph]].

\bibitem{Safdi:2025sfs}
B.~R.~Safdi and W.~L.~Xu,
``Wino and Real Minimal Dark Matter Excluded by Fermi Gamma-Ray Observations,''
[arXiv:2507.15934 [hep-ph]].

\bibitem{Cirelli:2007xd}
M.~Cirelli, A.~Strumia and M.~Tamburini,
``Cosmology and Astrophysics of Minimal Dark Matter,''
Nucl. Phys. B \textbf{787}, 152-175 (2007)
[arXiv:0706.4071 [hep-ph]].

\bibitem{Bottaro:2022one}
S.~Bottaro, D.~Buttazzo, M.~Costa, R.~Franceschini, P.~Panci, D.~Redigolo and L.~Vittorio,
``The last complex WIMPs standing,''
Eur. Phys. J. C \textbf{82}, no.11, 992 (2022)
[arXiv:2205.04486 [hep-ph]].

\bibitem{Wells:2003tf}
J.~D.~Wells,
``Implications of supersymmetry breaking with a little hierarchy between gauginos and scalars,''
[arXiv:hep-ph/0306127 [hep-ph]].

\bibitem{Arkani-Hamed:2004zhs}
N.~Arkani-Hamed, S.~Dimopoulos, G.~F.~Giudice and A.~Romanino,
``Aspects of split supersymmetry,''
Nucl. Phys. B \textbf{709}, 3-46 (2005)
[arXiv:hep-ph/0409232 [hep-ph]].

\bibitem{Wells:2004di}
J.~D.~Wells,
``PeV-scale supersymmetry,''
Phys. Rev. D \textbf{71}, 015013 (2005)
[arXiv:hep-ph/0411041 [hep-ph]].

\bibitem{Arvanitaki:2012ps}
A.~Arvanitaki, N.~Craig, S.~Dimopoulos and G.~Villadoro,
``Mini-Split,''
JHEP \textbf{02}, 126 (2013)
[arXiv:1210.0555 [hep-ph]].

\bibitem{Bhattiprolu:2023lfh}
P.~N.~Bhattiprolu and J.~D.~Wells,
``Precision unification and the scale of supersymmetry,''
Phys. Rev. D \textbf{109}, no.1, L011704 (2024)
[arXiv:2309.12954 [hep-ph]].

\bibitem{PandaX-4T:2021bab}
Y.~Meng \textit{et al.} [PandaX-4T],
``Dark Matter Search Results from the PandaX-4T Commissioning Run,''
Phys. Rev. Lett. \textbf{127}, no.26, 261802 (2021)
[arXiv:2107.13438 [hep-ex]].

\bibitem{XENON:2023cxc}
E.~Aprile \textit{et al.} [XENON],
``First Dark Matter Search with Nuclear Recoils from the XENONnT Experiment,''
Phys. Rev. Lett. \textbf{131}, no.4, 041003 (2023)
[arXiv:2303.14729 [hep-ex]].

\bibitem{LZ:2022lsv}
J.~Aalbers \textit{et al.} [LZ],
``First Dark Matter Search Results from the LUX-ZEPLIN (LZ) Experiment,''
Phys. Rev. Lett. \textbf{131}, no.4, 041002 (2023)
[arXiv:2207.03764 [hep-ex]].

\bibitem{PandaX:2024qfu}
Z.~Bo \textit{et al.} [PandaX],
``Dark Matter Search Results from 1.54{\,}{\,}Tonne{\textperiodcentered}Year Exposure of PandaX-4T,''
Phys. Rev. Lett. \textbf{134}, no.1, 011805 (2025)
[arXiv:2408.00664 [hep-ex]].

\bibitem{LZ:2024zvo}
J.~Aalbers \textit{et al.} [LZ],
``Dark Matter Search Results from 4.2{\,}{\,}Tonne-Years of Exposure of the LUX-ZEPLIN (LZ) Experiment,''
Phys. Rev. Lett. \textbf{135}, no.1, 011802 (2025)
[arXiv:2410.17036 [hep-ex]].

\bibitem{XLZD:2024nsu}
J.~Aalbers \textit{et al.} [XLZD],
``The XLZD Design Book: Towards the Next-Generation Liquid Xenon Observatory for Dark Matter and Neutrino Physics,''
[arXiv:2410.17137 [hep-ex]].

\bibitem{Baudis:2024jnk}
L.~Baudis,
``DARWIN/XLZD: A future xenon observatory for dark matter and other rare interactions,''
Nucl. Phys. B \textbf{1003}, 116473 (2024)
[arXiv:2404.19524 [astro-ph.IM]].

\bibitem{PANDA-X:2024dlo}
A.~Abdukerim \textit{et al.} [PANDA-X and PandaX],
``PandaX-xT\textemdash{}A deep underground multi-ten-tonne liquid xenon observatory,''
Sci. China Phys. Mech. Astron. \textbf{68}, no.2, 221011 (2025)
[arXiv:2402.03596 [hep-ex]].

\bibitem{Billard:2013qya}
J.~Billard, L.~Strigari and E.~Figueroa-Feliciano,
``Implication of neutrino backgrounds on the reach of next generation dark matter direct detection experiments,''
Phys. Rev. D \textbf{89}, no.2, 023524 (2014)
[arXiv:1307.5458 [hep-ph]].

\bibitem{Randall:1998uk}
L.~Randall and R.~Sundrum,
``Out of this world supersymmetry breaking,''
Nucl. Phys. B \textbf{557}, 79-118 (1999)
[arXiv:hep-th/9810155].

\bibitem{Giudice:1998xp}
G.~F.~Giudice, M.~A.~Luty, H.~Murayama and R.~Rattazzi,
``Gaugino mass without singlets,''
JHEP \textbf{12}, 027 (1998)
[arXiv:hep-ph/9810442].

\bibitem{Thomas:1998wy}
S.~D.~Thomas and J.~D.~Wells,
``Phenomenology of Massive Vectorlike Doublet Leptons,''
Phys. Rev. Lett. \textbf{81}, 34-37 (1998)
[arXiv:hep-ph/9804359 [hep-ph]].

\bibitem{Tucker-Smith:2001myb}
D.~Tucker-Smith and N.~Weiner,
``Inelastic dark matter,''
Phys. Rev. D \textbf{64}, 043502 (2001)
[arXiv:hep-ph/0101138 [hep-ph]].

\bibitem{Graham:2024syw}
P.~W.~Graham, H.~Ramani and S.~S.~Y.~Wong,
``Enhancing direct detection of Higgsino dark matter,''
Phys. Rev. D \textbf{111}, no.5, 055030 (2025)
[arXiv:2409.07768 [hep-ph]].

\bibitem{Griest:1990kh}
K.~Griest and D.~Seckel,
``Three exceptions in the calculation of relic abundances,''
Phys. Rev. D \textbf{43}, 3191-3203 (1991)

\bibitem{Bae:2013bva}
K.~J.~Bae, H.~Baer and E.~J.~Chun,
``Mainly axion cold dark matter from natural supersymmetry,''
Phys. Rev. D \textbf{89}, no.3, 031701 (2014)
[arXiv:1309.0519 [hep-ph]].

\bibitem{Bae:2014yta}
K.~J.~Bae, H.~Baer and H.~Serce,
``Natural little hierarchy for SUSY from radiative breaking of the Peccei-Quinn symmetry,''
Phys. Rev. D \textbf{91}, no.1, 015003 (2015)
[arXiv:1410.7500 [hep-ph]].

\bibitem{Bae:2017hlp}
K.~J.~Bae, H.~Baer and H.~Serce,
``Prospects for axion detection in natural SUSY with mixed axion-higgsino dark matter: back to invisible?,''
JCAP \textbf{06}, 024 (2017)
[arXiv:1705.01134 [hep-ph]].

\bibitem{Baer:2019uom}
H.~Baer, V.~Barger, D.~Sengupta, H.~Serce, K.~Sinha and R.~W.~Deal,
``Is the magnitude of the Peccei{\textendash}Quinn scale set by the landscape?,''
Eur. Phys. J. C \textbf{79}, no.11, 897 (2019)
[arXiv:1905.00443 [hep-ph]].

\bibitem{Rinchiuso:2020skh}
L.~Rinchiuso, O.~Macias, E.~Moulin, N.~L.~Rodd and T.~R.~Slatyer,
``Prospects for detecting heavy WIMP dark matter 
with the Cherenkov Telescope Array: The Wino and Higgsino,''
Phys. Rev. D \textbf{103}, no.2, 023011 (2021)
[arXiv:2008.00692 [astro-ph.HE]].

\bibitem{Rodd:2024qsi}
N.~L.~Rodd, B.~R.~Safdi and W.~L.~Xu,
``CTA and SWGO can discover Higgsino dark matter annihilation,''
Phys. Rev. D \textbf{110}, no.4, 043003 (2024)
[arXiv:2405.13104 [hep-ph]].

\bibitem{Abe:2025lci}
S.~Abe, T.~Inada, E.~Moulin, N.~L.~Rodd, B.~R.~Safdi and W.~L.~Xu,
``Discovering the Higgsino at CTAO-North within the Decade,''
[arXiv:2506.08084 [hep-ph]].

\bibitem{Baumgart:2025dov}
M.~Baumgart, S.~Bottaro, D.~Redigolo, N.~L.~Rodd and T.~R.~Slatyer,
``Testing Real WIMPs with CTAO,''
[arXiv:2507.15937 [hep-ph]].

\bibitem{CTAConsortium:2017dvg}
B.~S.~Acharya \textit{et al.} [CTA Consortium],
``Science with the Cherenkov Telescope Array,''
[arXiv:1709.07997 [astro-ph.IM]].

\bibitem{Albert:2019afb}
A.~Albert, R.~Alfaro, H.~Ashkar, C.~Alvarez, J.~Alvarez, J.~C.~Arteaga-Vel{\'a}zquez, H.~A.~Ayala Solares, R.~Arceo, J.~A.~Bellido and S.~BenZvi, \textit{et al.}
``Science Case for a Wide Field-of-View Very-High-Energy Gamma-Ray Observatory in the Southern Hemisphere,''
[arXiv:1902.08429 [astro-ph.HE]].

\bibitem{Martin:2024pxx}
S.~P.~Martin,
``Implications of purity constraints on light Higgsinos,''
Phys. Rev. D \textbf{109}, no.9, 095045 (2024)
[arXiv:2403.19598 [hep-ph]].

\bibitem{Martin:2024ytt}
S.~P.~Martin,
``Curtain lowers on directly detectable higgsino dark matter,''
Phys. Rev. D \textbf{111}, no.7, 075004 (2025)
[arXiv:2412.08958 [hep-ph]].

\bibitem{Belanger:2001fz}
G.~Belanger, F.~Boudjema, A.~Pukhov and A.~Semenov,
``MicrOMEGAs: A Program for calculating the relic density in the MSSM,''
Comput. Phys. Commun. \textbf{149}, 103-120 (2002)
[arXiv:hep-ph/0112278].

\bibitem{Belanger:2004yn}
G.~Belanger, F.~Boudjema, A.~Pukhov and A.~Semenov,
``micrOMEGAs: Version 1.3,''
Comput. Phys. Commun. \textbf{174}, 577-604 (2006)
[arXiv:hep-ph/0405253].

\bibitem{Belanger:2020gnr}
G.~Belanger, A.~Mjallal and A.~Pukhov,
``Recasting direct detection limits within micrOMEGAs and implication for 
non-standard Dark Matter scenarios,''
Eur. Phys. J. C \textbf{81}, no.3, 239 (2021)
[arXiv:2003.08621 [hep-ph]].

\bibitem{Alguero:2023zol}
G.~Alguero, G.~Belanger, F.~Boudjema, S.~Chakraborti, A.~Goudelis, S.~Kraml, 
A.~Mjallal and A.~Pukhov,
``micrOMEGAs 6.0: N-component dark matter,''
Comput. Phys. Commun. \textbf{299}, 109133 (2024)
[arXiv:2312.14894 [hep-ph]].

\bibitem{Djouadi:2002ze}
A.~Djouadi, J.~L.~Kneur and G.~Moultaka,
``SuSpect: A Fortran code for the supersymmetric and Higgs 
particle spectrum in the MSSM,''
Comput. Phys. Commun. \textbf{176}, 426-455 (2007)
[arXiv:hep-ph/0211331 [hep-ph]].

\bibitem{Kneur:2022vwt}
J.~L.~Kneur, G.~Moultaka, M.~Ughetto, D.~Zerwas and A.~Djouadi,
``SuSpect3: A C++ code for the supersymmetric and Higgs particle 
spectrum of the MSSM,''
Comput. Phys. Commun. \textbf{291}, 108805 (2023)
[arXiv:2211.16956 [hep-ph]].

\bibitem{Allanach:2001kg}
B.~C.~Allanach,
``SOFTSUSY: a program for calculating supersymmetric spectra,''
Comput. Phys. Commun. \textbf{143}, 305-331 (2002)
[arXiv:hep-ph/0104145].

\bibitem{OHare:2021utq} C.~A.~J.~O'Hare,
``New Definition of the Neutrino Floor for Direct Dark Matter Searches,''
Phys. Rev. Lett. \textbf{127}, no.25, 251802 (2021)
[arXiv:2109.03116 [hep-ph]].

\bibitem{NeutrinoFoggithub} C.~A.~J.~O'Hare,
{https://github.com/cajohare/NeutrinoFog}.

\bibitem{ATLAS:2022rme}
G.~Aad \textit{et al.} [ATLAS],
``Search for long-lived charginos based on a disappearing-track signature using 136 fb$^{-1}$ of pp collisions at $\sqrt{s}$~=~13~TeV with the ATLAS detector,''
Eur. Phys. J. C \textbf{82}, no.7, 606 (2022)
[arXiv:2201.02472 [hep-ex]].

\bibitem{CMS:2023mny}
A.~Hayrapetyan \textit{et al.} [CMS],
``Search for supersymmetry in final states with disappearing tracks in proton-proton collisions at s=13{\,}{\,}TeV,''
Phys. Rev. D \textbf{109}, no.7, 072007 (2024)
[arXiv:2309.16823 [hep-ex]].


\bibitem{ATLAS:2019lng}
G.~Aad \textit{et al.} [ATLAS],
``Searches for electroweak production of supersymmetric particles with compressed mass spectra in $\sqrt{s}=$ 13 TeV $pp$ collisions with the ATLAS detector,''
Phys. Rev. D \textbf{101}, no.5, 052005 (2020)
[arXiv:1911.12606 [hep-ex]].

\bibitem{ATLAS:2021moa}
G.~Aad \textit{et al.} [ATLAS],
``Search for chargino{\textendash}neutralino pair production in final states with three leptons and missing transverse momentum in $\sqrt{s} = 13$~TeV pp collisions with the ATLAS detector,''
Eur. Phys. J. C \textbf{81}, no.12, 1118 (2021)
[arXiv:2106.01676 [hep-ex]].

\bibitem{CMS:2021edw}
A.~Tumasyan \textit{et al.} [CMS],
``Search for supersymmetry in final states with two or three soft leptons and missing transverse momentum in proton-proton collisions at $ \sqrt{s} $ = 13 TeV,''
JHEP \textbf{04}, 091 (2022)
[arXiv:2111.06296 [hep-ex]].

\bibitem{CMS:2024gyw}
A.~Hayrapetyan \textit{et al.} [CMS],
``Combined search for electroweak production of winos, binos, higgsinos, and sleptons in proton-proton collisions at s=13{\,}{\,}TeV,''
Phys. Rev. D \textbf{109}, no.11, 112001 (2024)
[arXiv:2402.01888 [hep-ex]].


\bibitem{ATLAS:2024umc}
G.~Aad \textit{et al.} [ATLAS],
``Search for Nearly Mass-Degenerate Higgsinos Using Low-Momentum Mildly Displaced Tracks in pp Collisions at s=13\,\,TeV with the ATLAS Detector,''
Phys. Rev. Lett. \textbf{132}, no.22, 221801 (2024)
[arXiv:2401.14046 [hep-ex]].

\bibitem{CMSdisplacedsoft} 
CMS collaboration,
``Search for compressed electroweakinos with low-momentum isolated tracks,''
CMS-PAS-SUS-24-012, April 7, 2025,
https://cds.cern.ch/record/2929520


\bibitem{CMS:2025mie}
A.~Hayrapetyan \textit{et al.} [CMS],
``Search for Higgsinos in final states with low-momentum lepton-track
pairs at 13 TeV,''
[arXiv:2511.16394 [hep-ex]].

\bibitem{ATLAS:2025lhc}
G.~Aad \textit{et al.} [ATLAS],
``Search for higgsinos in compressed mass spectra using low-momentum
tracks in $pp$ collisions at $\sqrt{s}=13$ TeV with the ATLAS
detector,''
[arXiv:2511.20042 [hep-ex]].


\bibitem{ATLAS:2021kxv}
G.~Aad \textit{et al.} [ATLAS],
``Search for new phenomena in events with an energetic jet and missing transverse momentum in $pp$ collisions at $\sqrt {s}$ =13  TeV with the ATLAS detector,''
Phys. Rev. D \textbf{103}, no.11, 112006 (2021)
[arXiv:2102.10874 [hep-ex]].

\bibitem{CMS:2021far}
A.~Tumasyan \textit{et al.} [CMS],
``Search for new particles in events with energetic jets and large missing transverse momentum in proton-proton collisions at $ \sqrt{s} $ = 13 TeV,''
JHEP \textbf{11}, 153 (2021)
[arXiv:2107.13021 [hep-ex]].


\bibitem{Agin:2023yoq}
D.~Agin, B.~Fuks, M.~D.~Goodsell and T.~Murphy,
``Monojets reveal overlapping excesses for light compressed higgsinos,''
Phys. Lett. B \textbf{853}, 138597 (2024)
[arXiv:2311.17149 [hep-ph]].

\bibitem{Agin:2024yfs}
D.~Agin, B.~Fuks, M.~D.~Goodsell and T.~Murphy,
``Seeking a coherent explanation of LHC excesses for compressed spectra,''
Eur. Phys. J. C \textbf{84}, no.11, 1218 (2024)
[arXiv:2404.12423 [hep-ph]].

\bibitem{Agin:2025vgn}
D.~Agin, B.~Fuks, M.~D.~Goodsell and T.~Murphy,
``A joint explanation for the soft lepton and monojet LHC excesses in the wino-bino model,''
Eur. Phys. J. C \textbf{85}, no.10, 1145 (2025)
doi:10.1140/epjc/s10052-025-14886-4
[arXiv:2506.21676 [hep-ph]].

\bibitem{Araz:2025bww}
J.~Y.~Araz, B.~Fuks, M.~D.~Goodsell and T.~Murphy,
``Deciphering compressed electroweakino excesses with MadAnalysis 5,''
[arXiv:2507.08927 [hep-ph]].

\bibitem{Chakraborti:2024pdn}
M.~Chakraborti, S.~Heinemeyer and I.~Saha,
``Consistent excesses in the search for $\tilde{\chi }_{2}^0\tilde{\chi }_{1}^\pm :$ wino/bino vs. Higgsino dark matter,''
Eur. Phys. J. C \textbf{84}, no.8, 812 (2024)
[arXiv:2403.14759 [hep-ph]].

\bibitem{Ellwanger:2024vvs}
U.~Ellwanger, C.~Hugonie, S.~F.~King and S.~Moretti,
``NMSSM explanation for excesses in the search for neutralinos and charginos and a 95 GeV Higgs boson,''
Eur. Phys. J. C \textbf{84}, no.8, 788 (2024)
[arXiv:2404.19338 [hep-ph]].

\bibitem{Canepa:2020ntc}
A.~Canepa, T.~Han and X.~Wang,
``The Search for Electroweakinos,''
Ann. Rev. Nucl. Part. Sci. \textbf{70}, 425-454 (2020)
[arXiv:2003.05450 [hep-ph]].

\bibitem{Capdevilla:2024bwt}
R.~Capdevilla, F.~Meloni and J.~Zurita,
``Discovering Electroweak Interacting Dark Matter at Muon Colliders Using Soft Tracks,''
Phys. Rev. Lett. \textbf{134}, no.18, 181802 (2025)
[arXiv:2405.08858 [hep-ph]].


\bibitem{Chen:1996ap}
C.~H.~Chen, M.~Drees and J.~F.~Gunion,
``A Nonstandard string / SUSY scenario and its phenomenological implications,''
Phys. Rev. D \textbf{55}, 330-347 (1997)
[erratum: Phys. Rev. D \textbf{60}, 039901 (1999)]
[arXiv:hep-ph/9607421 [hep-ph]].

\bibitem{Ibe:2023dcu}
M.~Ibe, Y.~Nakayama and S.~Shirai,
``Precise estimate of charged Higgsino/Wino decay rate,''
JHEP \textbf{03}, 012 (2024)
[arXiv:2312.08087 [hep-ph]].


\bibitem{ATLAS:2017avc}
M.~Aaboud \textit{et al.} [ATLAS],
``Search for supersymmetry in events with $b$-tagged jets and missing transverse momentum in $pp$ collisions at $\sqrt{s}=13$ TeV with the ATLAS detector,''
JHEP \textbf{11}, 195 (2017)
[arXiv:1708.09266 [hep-ex]].

\bibitem{ATLAS:2021yij}
G.~Aad \textit{et al.} [ATLAS],
``Search for new phenomena in final states with $b$-jets and missing transverse momentum in $\sqrt{s}=13$ TeV $pp$ collisions with the ATLAS detector,''
JHEP \textbf{05}, 093 (2021)
[arXiv:2101.12527 [hep-ex]].

\bibitem{CMS:2019zmd}
CMS~Collaboration \textit{et al.} [CMS],
``Search for supersymmetry in proton-proton collisions at 13 TeV in final states with jets and missing transverse momentum,''
JHEP \textbf{10}, 244 (2019)
[arXiv:1908.04722 [hep-ex]].

\bibitem{Baer:2023uwo}
H.~Baer, V.~Barger, J.~Dutta, D.~Sengupta and K.~Zhang,
``Top squarks from the landscape at high luminosity LHC,''
Phys. Rev. D \textbf{108}, no.7, 075027 (2023)
[arXiv:2307.08067 [hep-ph]].


\bibitem{Baer:2013yha}
H.~Baer, V.~Barger, P.~Huang, D.~Mickelson, A.~Mustafayev, W.~Sreethawong and X.~Tata,
``Same sign diboson signature from supersymmetry models with light higgsinos at the LHC,''
Phys. Rev. Lett. \textbf{110}, no.15, 151801 (2013)
[arXiv:1302.5816 [hep-ph]].

\bibitem{Baer:2023olq}
H.~Baer, V.~Barger, X.~Tata and K.~Zhang,
``Winos from natural SUSY at the high luminosity LHC,''
Phys. Rev. D \textbf{109}, no.1, 015027 (2024)
[arXiv:2310.10829 [hep-ph]].

\bibitem{Baer:2023ech}
H.~Baer, V.~Barger, J.~Bolich, J.~Dutta and D.~Sengupta,
``Natural anomaly mediation from the landscape with implications for LHC SUSY searches,''
Phys. Rev. D \textbf{109}, no.3, 035011 (2024)
[arXiv:2311.18120 [hep-ph]].

\bibitem{Co:2022jsn}
R.~T.~Co, A.~Pierce, B.~Sheff and J.~D.~Wells,
``Discovery potential for split supersymmetry with thermal dark matter,''
Phys. Rev. D \textbf{106}, no.9, 095001 (2022)
[arXiv:2206.11912 [hep-ph]].

\end{thebibliography}
\end{document}